%% file: manuscript.tex
\documentclass[aps,prd,reprint,superscriptaddress,nofootinbib]{revtex4-2}

\usepackage{amsmath,amssymb,bm}
\usepackage{graphicx}
\usepackage{booktabs}
\usepackage{xcolor}
\usepackage[hidelinks]{hyperref}

\newcommand{\efo}{\epsilon_{2,\mathrm{fo}}}
\newcommand{\ein}{\epsilon_{2,\mathrm{init}}}
\newcommand{\Pz}{P_{z,s2}}
\newcommand{\Cdet}{C_{\mathrm{det}}}

\begin{document}

\title{Longitudinal \texorpdfstring{$\Lambda$}{Lambda} polarization as a
competition between freeze-out geometry and elliptic flow in Pb-Pb collisions at
\texorpdfstring{$\sqrt{s_{\mathrm{NN}}}=5.36\,\mathrm{TeV}$}{sqrt(sNN)=5.36 TeV}}

\author{Yilong Xie}
\email{xieyl@cug.edu.cn}
\affiliation{School of Mathematics and Physics, China University of Geosciences (Wuhan), Lumo Road 388, 430074 Wuhan, China}
\author{Simin Wu}
\affiliation{School of Mathematics and Physics, China University of Geosciences (Wuhan), Lumo Road 388, 430074 Wuhan, China}
\author{Laszlo P. Csernai}
\affiliation{Institute of Physics and Technology, University of Bergen, Allegaten 55, 5007 Bergen, Norway}

\begin{abstract}
The second sine harmonic $\Pz$ of longitudinal $\Lambda$ polarization is
sensitive to the balance between the kinematic-vorticity contribution
$\Pz^\omega$ and the kinematic-shear contribution $\Pz^\Xi$ at decoupling.
We quantify this competition in a scan of 48 Pb-Pb initial states evolved
with the (3+1)-dimensional Particle-In-Cell Relativistic (PICR) ideal-hydrodynamic
code. With the isothermal
local-equilibrium (ILE) prescription, the total harmonic follows
$\Pz\simeq a\,\efo-b\,v_2$, with $a\simeq0.010$ and $b\simeq0.014$,
where $\efo$ is the freeze-out eccentricity and $v_2$ the elliptic flow of
$\Lambda$ hyperons. With coefficients fixed on these 48 states, the relation
describes 86\% of the variation of $\Pz$ in 12 states outside the fit. To identify the origin of the two terms, we decompose the ILE integrand exactly into an emission-gradient term and a curl term. On the PICR decoupling surfaces, the tangential emission gradient is
positive on average, correlates most strongly with $\efo$, and admits a surface-curvature
representation, whereas the normal emission gradient and the curl term are
negative on average and correlate most strongly with $v_2$. A blast-wave calculation with the same
integrand gives coefficients of comparable magnitude and shows how the
cancellation point $\efo/v_2=b/a$ changes with the flow and the surface geometry; both PICR and blast-wave calculations predict that $b/a$ rises with hadron mass from the proton to the $\Omega$ when the decoupling surface is held fixed. Although the geometry inferred from ALICE data depends on the background, the measured polarization constrains the displacement of $\efo$ from the cancellation boundary $\efo=(b/a)v_2$. For the central ALICE measurement values at $5.36\,\mathrm{TeV}$, the backgrounds examined give $\efo\simeq0.2$--$0.4$ in 30--50\% Pb-Pb collisions.
\end{abstract}

\maketitle

\section{Introduction}\label{sec:intro}

Relativistic hydrodynamics maps the initial geometry of a heavy-ion collision
onto final-state spectra and anisotropic flow~\cite{heinz2013,gale2013}. Spin polarization of the emitted hadrons probes the local gradients of the fluid. In noncentral collisions, the
large orbital angular momentum of the colliding nuclei was predicted to
polarize the produced hadrons~\cite{liangwang2005,becattini2008}. In the thermal-vorticity description of local thermodynamic equilibrium, the mean spin vector of an emitted particle depends on the antisymmetric gradient of the inverse-temperature four-vector $u^\mu/T$ at emission~\cite{becattini2013}. Polarization therefore probes the velocity and temperature gradients on the decoupling surface~\cite{becattini2020,huang2021,niida2024}. The global $\Lambda$
polarization measured by the STAR Collaboration~\cite{star2017,star2018}
established the coupling between spin and the rotation of the produced matter;
global polarization has also been investigated at the LHC by the ALICE
Collaboration, whose values are consistent with zero~\cite{alice2020}, and
measured at SIS18 energies by the HADES Collaboration~\cite{hades2022}. Hydrodynamic and transport calculations based on the
thermal vorticity describe the magnitude of global $\Lambda$ polarization and its decrease with collision energy~\cite{becattini2015b,karpenko2017,li2017,sun2017,wei2019,jiang2016,denghuang2016}.

The longitudinal component $P_z$ of the polarization depends on the azimuthal
angle $\phi$ of the hyperon momentum and therefore retains local information
that disappears in the global average. Its second sine harmonic $\Pz$ was
predicted from the thermal vorticity with the sign opposite to the
measurements~\cite{becattini2018,xia2018} at RHIC and the
LHC~\cite{star2019,alice2022,star2023}. This sign problem motivated two lines
of work. One treats spin as a dynamical degree of freedom in spin
hydrodynamics~\cite{ffjs2018,hidaka2022}. The other adds to the
local-equilibrium spin vector a contribution from the symmetric gradient of
$u^\mu/T$, called thermal shear, or from the symmetric gradient of $u^\mu$
in the isothermal prescription, called kinematic shear~\cite{bbp2021a,liuyin2021,buzzegoli2022}. The positive kinematic-shear contribution can outweigh the negative kinematic-vorticity contribution and reproduce the measured sign~\cite{bbp2021b,fu2021,fuxu2021,yi2021,florkowski2022,palermo2024}.
The outcome of this competition is sensitive to the spin
prescription~\cite{fukushima2026,palermo2024}, to the initial conditions and
the longitudinal structure of the source~\cite{wu2026initial,ryu2021,alzhrani2022},
to shear and bulk viscosity~\cite{palermo2025,singh2026}, and to feed-down
from heavier resonances~\cite{bcs2019,xia2019}.

The physical origin of the sign is known qualitatively. In blast-wave estimates
the longitudinal vorticity is proportional to the difference between the flow
anisotropy and the spatial anisotropy of the source~\cite{voloshin2017,niida2024}.
For a longitudinally boost-invariant ideal uncharged fluid with vanishing
initial transverse flow, the thermal vorticity reduces to an acceleration term,
and on a decoupling surface of constant proper time with $T=T(\tau)$, $P_z$ is
proportional to the cooling rate times $v_2$~\cite{becattini2018}. An analytically solvable blast-wave model with
flow-momentum correspondence attributes the polarization to the ellipticities
of the flow and of the emission area and identifies a symmetry between the
vorticity and shear contributions~\cite{arslan2024}. In anisotropic Gubser flow,
the time-derivative terms of the laboratory-frame prescription cancel between
the vorticity and shear contributions~\cite{fukushima2026}.

The preceding results relate longitudinal polarization to spatial and flow
anisotropies in specified backgrounds, i.e., decoupling surfaces together with
the flow and temperature fields in their neighborhood. To connect these mechanisms to a
three-dimensional hydrodynamic calculation, we examine whether the total harmonic $\Pz$
can be expressed through the freeze-out eccentricity $\efo$, which measures the
spatial anisotropy retained at emission, and the elliptic flow $v_2$, which
measures the momentum anisotropy. We determine the relation on a set of initial
states and test its accuracy on states outside that calibration set.

Section~\ref{sec:empirical} establishes $\Pz\simeq a\efo-bv_2$ on 48
ideal-hydrodynamic states and validates it on 12 states outside the fit.
Section~\ref{sec:origin} explains the competition between the geometric and
flow terms through exact identities of the ILE integrand, a heavy-hadron
saddle-point approximation, and the computed decoupling surfaces. The blast-wave
calculation of Sec.~\ref{sec:bw} then examines how the coefficient magnitudes
and the cancellation point $b/a$ change with the freeze-out background and the
hadron mass, and Sec.~\ref{sec:scope} infers freeze-out geometry from ALICE
data and evaluates how that inference depends on the background. The
Appendices give the derivations and numerical methods; the Supplemental
Material~\cite{supp} contains the state tables and additional validation. An
earlier version reported the compact empirical relation and its
validation~\cite{xie2026prl}; the present work develops the linear relation and
its interpretation in terms of velocity gradients and freeze-out geometry. The
linear-response construction of such relations and their extension to the
third harmonic are studied separately~\cite{xie2026resp,xie2026hier}.

\section{Model, polarization prescription, and observables}\label{sec:model}

\paragraph{Initial states.} We use Particle-In-Cell Relativistic (PICR) ideal
hydrodynamics for Pb-Pb collisions at
$\sqrt{s_{\mathrm{NN}}}=5.36\,\mathrm{TeV}$~\cite{picr2017}, with the
initial-state construction of Ref.~\cite{wu2026initial}. The initial energy
density on the Cartesian time slice $t_0=4\,\mathrm{fm}/c$ is
\begin{equation}
e(x,y,\eta_s)=e_{\mathrm{peak}}\,W_\perp(x,y)\,H(\eta_s).
\label{eq:init}
\end{equation}
The transverse weight $W_\perp$ is the Magas-Csernai-Strottman participant
weight~\cite{mcs2001}, normalized to its maximum and convolved with a
two-dimensional Gaussian of transverse smoothing width $\sigma_\perp$; the
normalization of the convolution preserves the transverse integral. The
initial energy-density normalization $e_{\mathrm{peak}}$ sets the scale before
smoothing. The reduced impact parameter is $b_0=b/(R_P+R_T)$, where $R_P$ and
$R_T$ are the nuclear radii; the geometrical estimate $c\simeq b_0^2$
associates $b_0=0.60$ and $b_0=0.77$ with centralities of about 36\% and 59\%.
The longitudinal profile $H(\eta_s)$ has a central plateau of half-width $0.8$
and Gaussian tails of width $0.6$ in space-time rapidity
$\eta_s=\tfrac12\ln[(t+z)/(t-z)]$. The initial longitudinal flow is the
Bjorken flow~\cite{bjorken}, $y_f=\eta_s$, where $y_f$ is the longitudinal
flow rapidity; the initial transverse velocity vanishes. The resulting initial state is reflection
symmetric under $z\to-z$ and carries no angular momentum, $J_y=0$; with this
fixed longitudinal background, the scan isolates the effect of the initial
transverse geometry. The $x$
axis lies along the impact parameter. For these smooth initial states the
second-order event plane coincides with the reaction plane, $\Psi_2=0$.

The scan contains 48 initial states, which combine four initial energy-density
normalizations $e_{\mathrm{peak}}=10$, $20$, $30$,
$40\,\mathrm{GeV}/\mathrm{fm}^3$, two reduced impact parameters $b_0=0.60$ and
$0.77$, and six transverse smoothing widths $\sigma_\perp=0.8$, $1.0$, $1.1$,
$1.2$, $1.35$, and $1.5\,\mathrm{fm}$. Increasing the transverse smoothing width $\sigma_\perp$ broadens the initial energy-density profile and weakens its local gradients; increasing the initial energy-density normalization $e_{\mathrm{peak}}$ lengthens the evolution. Twelve further states, which
do not enter any fit, test the relations of Sec.~\ref{sec:empirical}.

\paragraph{Evolution and decoupling.} The fluid obeys
$\partial_\mu T^{\mu\nu}=0$ with $T^{\mu\nu}=(e+p)u^\mu u^\nu-pg^{\mu\nu}$ and
the three-flavor bag-model equation of state of PICR~\cite{bagmit}, $p=(e-4B)/3$
in the plasma phase. At the decoupling temperature
$T_{\mathrm{dec}}=160\,\mathrm{MeV}$ this equation of state gives $p/e=0.027$,
compared with $0.17$--$0.20$ from lattice QCD~\cite{hotqcd}. Although the resulting pressure gradients and decoupling surfaces limit
quantitative comparisons with data in Sec.~\ref{sec:scope}, keeping this
equation of state fixed allows the scan to resolve how changes of the initial
transverse profile modify the competition between the kinematic-vorticity and
kinematic-shear contributions. The
isothermal decoupling surface $\Sigma$, defined by $T=T_{\mathrm{dec}}$, is
constructed with Cornelius~\cite{cornelius}.

\paragraph{Polarization.} The polarization of directly emitted $\Lambda$ and
$\bar\Lambda$ hyperons is computed with the laboratory-frame ILE prescription of
Becattini, Buzzegoli, Palermo, Inghirami, and
Karpenko~\cite{bbp2021a,bbp2021b}. With $\hbar=c=1$, the metric
$g=\mathrm{diag}(1,-1,-1,-1)$, $\epsilon^{0123}=+1$, and the laboratory time
direction $\hat t=(1,0,0,0)$, the polarization four-vector is
\begin{equation}
P^\mu(p)=-\frac{1}{2mT}\left\langle
\tfrac12\epsilon^{\mu\nu\alpha\beta}p_\beta\,\omega_{\alpha\nu}
+\frac{\epsilon^{\mu0\alpha\beta}p_\beta\,\Xi_{\alpha\lambda}p^\lambda}{E}
\right\rangle_\Sigma ,
\label{eq:ile}
\end{equation}
where $\omega_{\alpha\nu}=\tfrac12(\partial_\alpha u_\nu-\partial_\nu u_\alpha)$
is the kinematic vorticity,
$\Xi_{\alpha\nu}=\tfrac12(\partial_\alpha u_\nu+\partial_\nu u_\alpha)$ the
kinematic shear, and $E=p^0$. The surface average is
$\langle A\rangle_\Sigma=\int_\Sigma(p\cdot d\Sigma)\,f\,A/\int_\Sigma(p\cdot d\Sigma)\,f$
with $f=e^{-p\cdot u/T}$ (Boltzmann statistics, zero chemical potential); this
is the Cooper-Frye weight~\cite{cooperfrye}, and the sign of $p\cdot d\Sigma$
is retained. The component along the beam axis in
the $\Lambda$ rest frame is $P^*_z=P^z-p^zP^0/(E+m)$. The first term of
Eq.~(\ref{eq:ile}) gives the kinematic-vorticity contribution and the second
term the kinematic-shear contribution. Writing Eq.~(\ref{eq:ile}) as
$P^\mu=-\langle A^\mu\rangle_\Sigma/(2mT)$, we define the ILE integrand
\begin{equation}
K=-A^z+\frac{p^zA^0}{E+m},\qquad P^*_z=\frac{\langle K\rangle_\Sigma}{2mT};
\label{eq:Kdef}
\end{equation}
the analysis of Sec.~\ref{sec:origin} rests on the algebraic properties of
$K$.

\paragraph{Observables.} The harmonics are weighted with the measure
$d\Gamma=p_T\,dp_T\,d\phi\,F(p_T,\phi)$, where $F$ is the $\Lambda+\bar\Lambda$
yield averaged over the rapidity interval $|y|<0.5$. The elliptic flow and the
polarization harmonic are
\begin{align}
v_2&=\frac{\int d\Gamma\cos2(\phi-\Psi_2)}{\int d\Gamma},\nonumber\\
\Pz&=\frac{\int d\Gamma\,P^*_z\sin2(\phi-\Psi_2)}{\int d\Gamma},
\label{eq:harm}
\end{align}
integrated over $0.5<p_T<6\,\mathrm{GeV}/c$, where $P^*_z(p_T,\phi)$ is
averaged over rapidity with the same yield weight. The sine moment in
Eq.~(\ref{eq:harm}) defines $\Pz$ without an additional factor of two. The kinematic-vorticity
contribution $\Pz^\omega$ and the kinematic-shear contribution $\Pz^\Xi$ add to
the total ILE harmonic $\Pz^{\mathrm{ILE}}\equiv\Pz$. Their competition is
measured by the ratio $R=|\Pz^\Xi/\Pz^\omega|$, and goodness of fit is quoted as
the coefficient of determination $\Cdet$.

To test which anisotropy controls each polarization contribution, we compare
$\Pz^\omega$ and $\Pz^\Xi$ with three anisotropies of different stages. The
initial eccentricity
\begin{equation}
\ein=\frac{\int dx\,dy\,W_\perp\,(y^2-x^2)}{\int dx\,dy\,W_\perp\,(y^2+x^2)}
\end{equation}
characterizes the spatial anisotropy at the start of the evolution. The
freeze-out eccentricity
\begin{equation}
\efo=
\frac{\int_\Sigma (u^\mu d\Sigma_\mu)(y^2-x^2)}
{\int_\Sigma (u^\mu d\Sigma_\mu)(y^2+x^2)}
\label{eq:epsfo}
\end{equation}
characterizes the spatial anisotropy that remains at emission. The signed
fluid-flux weight $u^\mu d\Sigma_\mu$ is proportional to the
local-equilibrium particle flux on the isothermal surface; $\efo$ is therefore
a geometric quantity of the model, which Sec.~\ref{sec:scope} infers from the
measured polarization. The elliptic flow
$v_2$ characterizes the momentum anisotropy built up by the evolution. The
eccentricity survival fraction $S_\epsilon=\efo/\ein$ measures how much of the
initial spatial anisotropy remains at decoupling.

\section{Linear relation between \texorpdfstring{$\Pz$}{Pz,s2}, freeze-out eccentricity, and elliptic flow}\label{sec:empirical}

\subsection{Kinematic-vorticity and kinematic-shear contributions}

\begin{figure*}[t]
\centering
\includegraphics[width=0.82\textwidth]{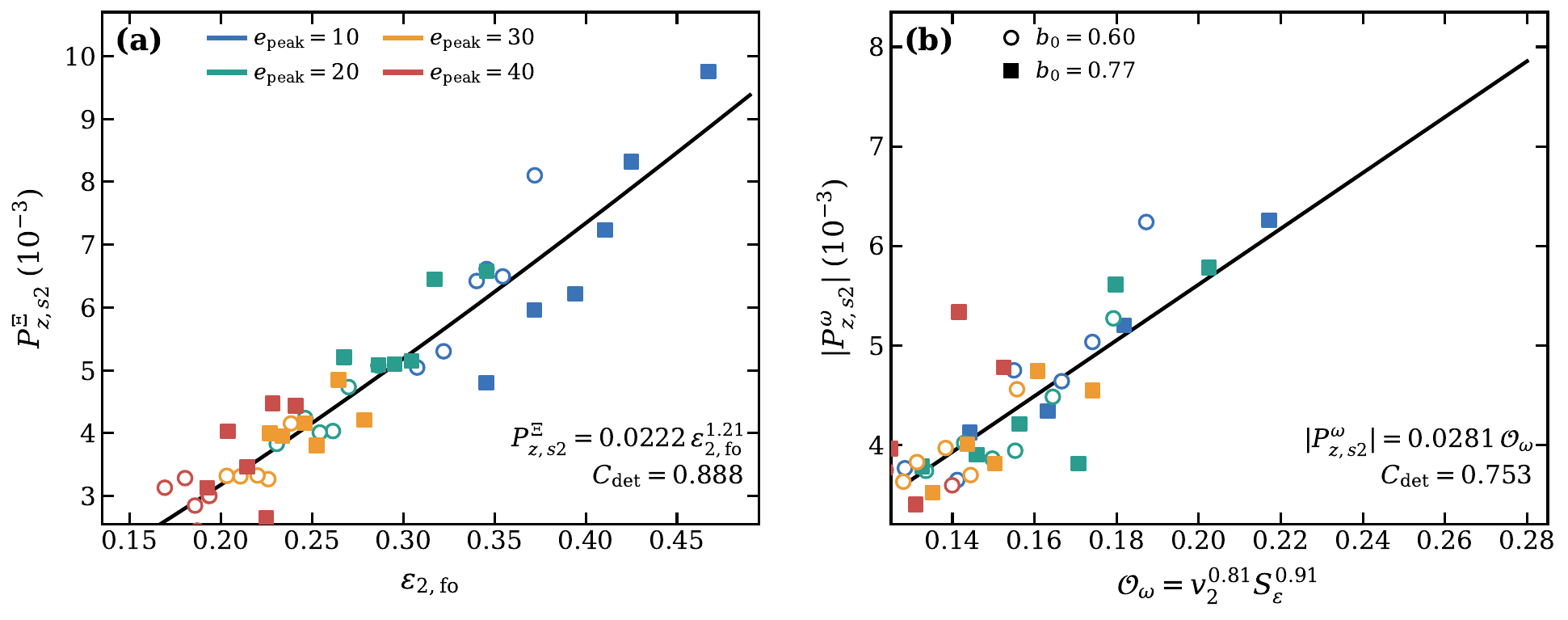}
\caption{Polarization contributions for the 48 initial states. (a)
Kinematic-shear contribution $\Pz^\Xi$ versus the freeze-out eccentricity
$\efo$. (b) Magnitude of the kinematic-vorticity contribution $|\Pz^\omega|$
versus $v_2^{0.81}S_\epsilon^{0.91}$. Colors denote the initial
energy-density normalization $e_{\mathrm{peak}}=10$, $20$, $30$, and
$40\,\mathrm{GeV}/\mathrm{fm}^3$; marker shapes denote the reduced impact
parameter $b_0=0.60$ and $0.77$; each combination contains the six transverse
smoothing widths $\sigma_\perp$. Solid curves are Eqs.~(\ref{eq:shear}) and
(\ref{eq:vort}).}
\label{fig:components}
\end{figure*}

Because the total harmonic $\Pz$ is a near cancellation, we first examine how
each of its two contributions depends on the anisotropies defined in
Sec.~\ref{sec:model}. Figure~\ref{fig:components}(a) shows $\Pz^\Xi$ as a
function of the freeze-out eccentricity $\efo$, and
Fig.~\ref{fig:components}(b) shows $|\Pz^\omega|$ as a function of the elliptic
flow $v_2$ combined with the survival fraction $S_\epsilon$. The kinematic-vorticity contribution is negative and the kinematic-shear
contribution positive in every scanned state. Over the 48 states they follow
\begin{align}
\Pz^{\Xi}&\simeq0.0222\,\efo^{1.21},
& \Cdet&=0.89,
\label{eq:shear}\\
\left|\Pz^{\omega}\right|&\simeq
0.0281\,v_2^{0.81}S_\epsilon^{0.91},
& \Cdet&=0.75.
\label{eq:vort}
\end{align}
By contrast, a fit of $\Pz^\Xi$ with the initial eccentricity $\ein$ alone
gives $\Cdet=0.16$, and a fit of $|\Pz^\omega|$ with $v_2$ alone gives
$\Cdet=0.17$. The kinematic-shear contribution $\Pz^\Xi$ therefore follows the
spatial anisotropy at emission rather than the initial one, and the
kinematic-vorticity contribution $\Pz^\omega$ follows the momentum anisotropy
$v_2$ modulated by $S_\epsilon$.

We next test whether the balance of the two contributions can be expressed
through the single ratio $\efo/v_2$. Dividing Eq.~(\ref{eq:shear}) by
Eq.~(\ref{eq:vort}) would still leave $S_\epsilon$, so a collapse onto
$\efo/v_2$ is an additional empirical regularity that the two power laws do not
imply.
Figure~\ref{fig:compact} (left) shows that the competition ratio
$R=|\Pz^\Xi/\Pz^\omega|$ indeed collapses onto a common trend in $\efo/v_2$,
\begin{equation}
R\simeq0.84\left(\efo/v_2\right)^{0.47},\qquad \Cdet=0.81.
\label{eq:ratio}
\end{equation}
The value of $\efo/v_2$ at which $R=1$ locates the cancellation of $\Pz^\omega$ and $\Pz^\Xi$. Predicting the magnitude of their sum also requires a relation for the total harmonic.

\subsection{Compact and linear relations for the total harmonic}

To include the magnitude as well as the sign of $\Pz$, we use the exact identity $\Pz=|\Pz^{\omega}|(R-1)$ to guide a fit directly to the total harmonic. Fitting the two coefficients gives the compact relation
\begin{equation}
\Pz\simeq
0.0404\,v_2S_\epsilon
\left[0.850\left(\efo/v_2\right)^{1/2}-1\right],
\label{eq:total}
\end{equation}
which vanishes at $\efo/v_2=1.38$. Equation~(\ref{eq:total}) gives
$\Cdet=0.87$ on the 48 states and the correct sign for 43 of them. A
least-squares fit without intercept in the two variables $\efo$ and $v_2$ gives
the simpler linear relation
\begin{equation}
\Pz\simeq a\,\efo-b\,v_2,\qquad a=0.0100,\ b=0.0136,
\label{eq:linear}
\end{equation}
which vanishes at $\efo/v_2=b/a=1.36$. Table~\ref{tab:linear} compares the two
relations: they give nearly equal $\Cdet$ on the calibration states and equal values to
three decimal places on the 12
states outside the fit, and their predictions have a correlation coefficient of
$0.997$ on the 48 states. The two forms serve different purposes. The linear
relation is written in the final-state geometric and momentum anisotropies and
makes the coefficients and the cancellation point explicit. The compact
relation keeps the factorized structure of the identity $\Pz=|\Pz^{\omega}|(R-1)$ and contains the initial eccentricity
explicitly, which Sec.~\ref{sec:scope} uses to show how the inferred geometry
depends on the assumed initial geometry.

\begin{table}[t]
\caption{Three relations for the total harmonic $\Pz$. Fitted coefficients:
$0.0404$ and $0.850$ for Eq.~(\ref{eq:total}); $a=0.0100$ and $b=0.0136$ for
Eq.~(\ref{eq:linear}); $0.0155$ and $0.00886$ for $a\efo-b\,\ein$. The second
column gives the coefficient of determination $\Cdet$ on the 48 calibration
states and on the 12 additional states, the latter with coefficients fixed on
the calibration states. The third column gives the numbers of correct signs for the 48 and 12 states,
in that order. The final column gives pooled $\Cdet$ from grouped
cross-validation, withholding groups of equal $e_{\mathrm{peak}}$,
$\sigma_\perp$, and $b_0$, respectively.}
\label{tab:linear}
\centering\footnotesize
\setlength{\tabcolsep}{3pt}
\begin{tabular}{lccc}
\toprule
relation & $\Cdet$ (48/12) & signs & grouped CV\\
\midrule
Eq.~(\ref{eq:total}) & 0.865/0.860 & 43, 10 & 0.81, 0.84, 0.72\\
Eq.~(\ref{eq:linear}), $a\efo-bv_2$ & 0.866/0.860 & 41, 10 & 0.86, 0.83, 0.74\\
$a\efo-b\,\ein$ & 0.880/0.894 & 42, 11 & 0.86, 0.86, 0.72\\
\bottomrule
\end{tabular}
\end{table}

To understand why the compact and linear relations have nearly equal
accuracy, we examine the relations among the three anisotropies in the same
scan. They satisfy
\begin{equation}
\sqrt{\efo v_2}\simeq0.49\,\ein,\qquad
S_\epsilon v_2\simeq0.186\,\efo+0.313\,v_2,
\label{eq:conversion}
\end{equation}
which hold with a relative rms deviation of 6\% and with $\Cdet=0.86$,
respectively. The first regularity turns the square-root term of
Eq.~(\ref{eq:total}) into $0.0168\,\efo$, and the second converts the remaining
term $-0.0404\,S_\epsilon v_2$. Together they give
$\Pz\simeq0.0093\,\efo-0.0126\,v_2$, close to Eq.~(\ref{eq:linear}). Both
regularities accompany the near-planarity of the three anisotropies,
$\ein\simeq0.62\,\efo+1.54\,v_2$ (rms deviation 3.7\%), which also explains why
the form $a\,\efo-b\,\ein$ performs well (Table~\ref{tab:linear}). The agreement also extends to the cancellation point: the value
$\efo/v_2\simeq1.45$ at which the fit~(\ref{eq:ratio}) gives $R=1$ and the zeros
$1.38$ and $1.36$ of Eqs.~(\ref{eq:total}) and (\ref{eq:linear}) come from
separate fits and agree to within 7\%. Among these comparably accurate
empirical forms we use
Eq.~(\ref{eq:linear}) in
the analysis below, because the ILE prescription is local on the decoupling
surface and $(\efo,v_2)$ are both defined there (Sec.~\ref{sec:sym}). The
fitting procedure is given in Appendix~\ref{app:fit}.

\subsection{Validation on states outside the fit}

\begin{figure*}[t]
\centering
\includegraphics[width=0.46\textwidth]{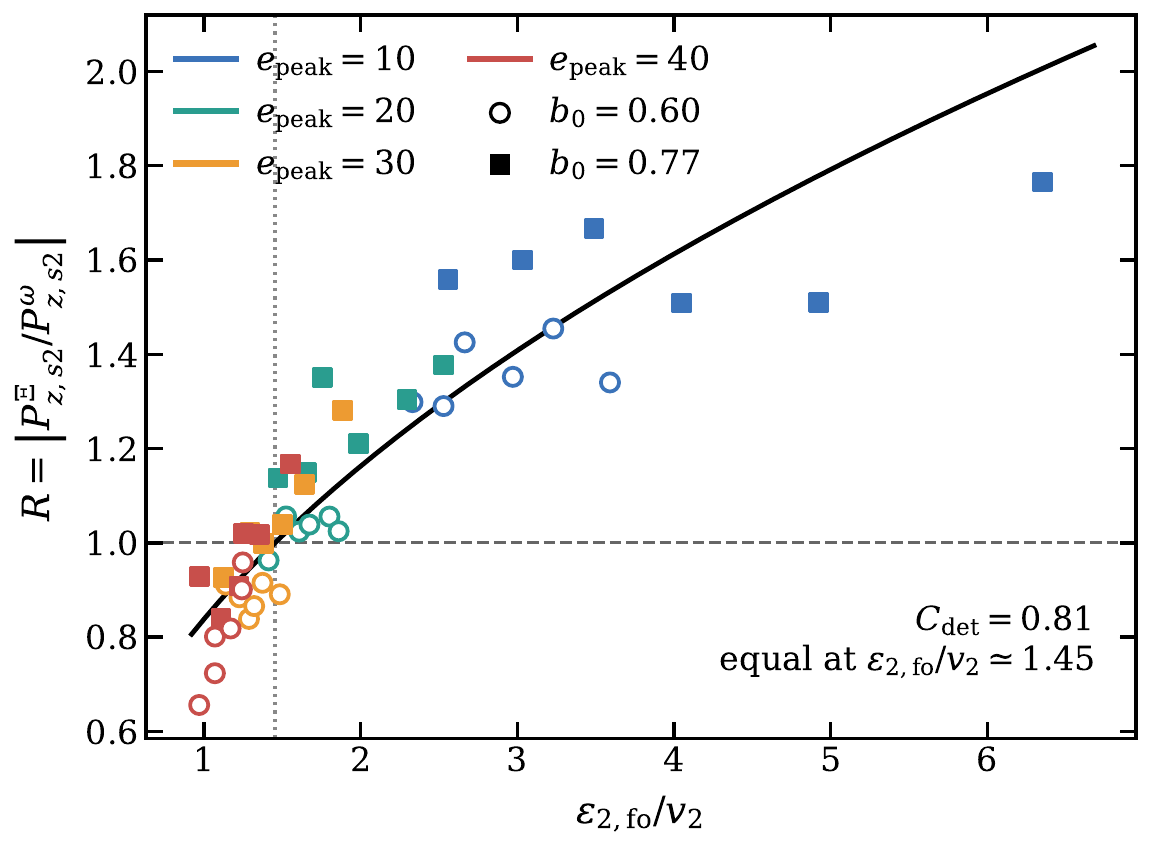}\hfill
\includegraphics[width=0.46\textwidth]{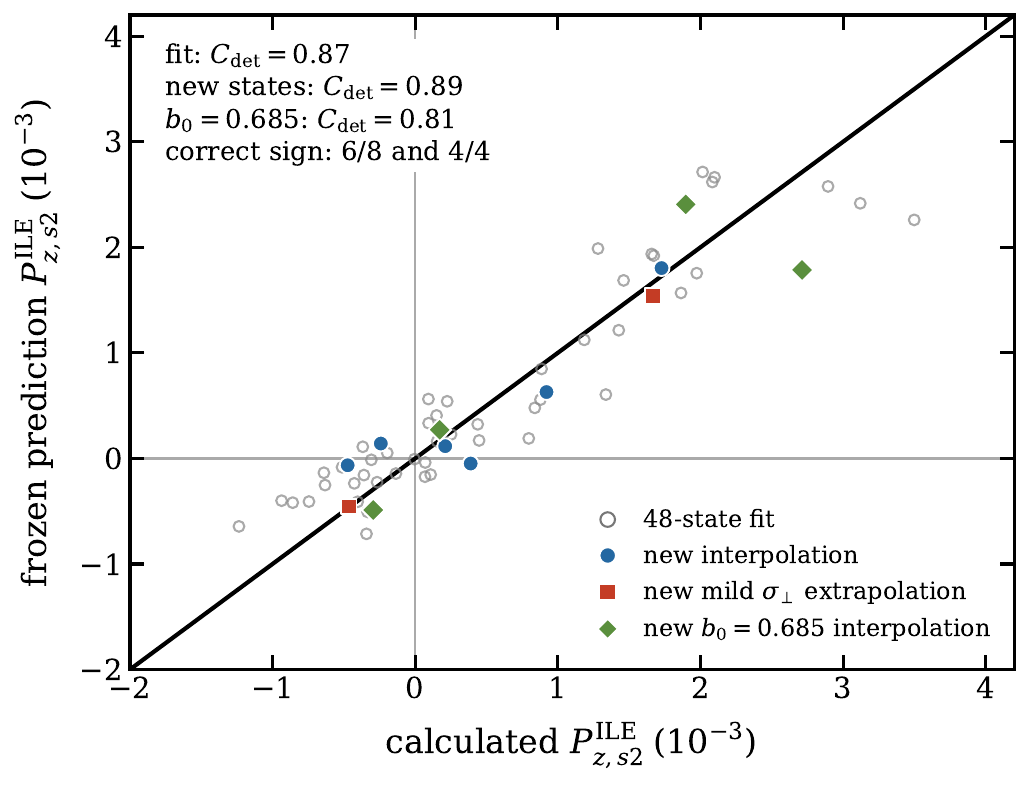}
\caption{Left: the competition ratio $R=|\Pz^\Xi/\Pz^\omega|$ versus
$\efo/v_2$ for the 48 initial states; the dashed line marks $R=1$. Right: total ILE harmonic $\Pz$ predicted by the compact relation~(\ref{eq:total}) versus the direct calculation. Open gray circles are the 48 calibration states; blue circles are
six additional states inside the scanned range, red squares two states with
$\sigma_\perp=0.7$ and $1.6\,\mathrm{fm}$, and green diamonds four states at
$b_0=0.685$. The predictions of the linear relation~(\ref{eq:linear}) nearly
coincide with those of Eq.~(\ref{eq:total}) (Supplemental Material). The coefficients are not refitted to any of the 12 additional
states.}
\label{fig:compact}
\end{figure*}

To test the relation beyond its calibration set, we use eight additional
states and four states at the previously unsampled reduced impact parameter
$b_0=0.685$. Six of the eight states interpolate within the scanned range;
the other two extend the transverse smoothing width to $\sigma_\perp=0.7$
and $1.6\,\mathrm{fm}$. None of the 12 states enters either fit. The compact
coefficients were fixed before these calculations, whereas the linear
relation was introduced later and fitted only on the 48 calibration states.
Figure~\ref{fig:compact} (right) compares the predictions of the compact
relation~(\ref{eq:total}) with the direct calculations; the corresponding tests of
the linear relation~(\ref{eq:linear}) are given in Table~\ref{tab:linear} and in
Fig.~S1 of the Supplemental Material~\cite{supp}.

For the eight additional states, Eq.~(\ref{eq:linear}) gives $\Cdet=0.89$
and a root-mean-square error of $0.28\times10^{-3}$. The two sign misses
have $|\Pz|<0.4\times10^{-3}$ and lie near the cancellation boundary
$\efo/v_2\simeq1.4$. For the four states at $b_0=0.685$, the relation gives
$\Cdet=0.81$ and all four signs correctly. Although the sign remains
sensitive to small absolute errors near cancellation, the fixed linear
relation predicts 10 of the 12 signs and describes 86\% of the variation of
$\Pz$. The compact relation gives the same coefficient of determination and
root-mean-square error to two significant figures. These tests support the
relation within and slightly beyond the scanned initial-state range.

Changing the decoupling temperature also changes the surface on which the
relation is evaluated. We therefore test the fixed coefficients at
$e_{\mathrm{peak}}=10\,\mathrm{GeV}/\mathrm{fm}^3$,
$\sigma_\perp=1.5\,\mathrm{fm}$, and $b_0=0.77$, at four temperatures
between 157.5 and
$170\,\mathrm{MeV}$ (Supplemental Material~\cite{supp}). The linear relation
has absolute deviations of $0.63$--$0.75\times10^{-3}$, two to three times its
root-mean-square error on the eight additional states. Despite these amplitude deviations, the fixed coefficients reproduce the positive, shear-dominated sign at all four temperatures for this initial state.

\section{Origin of the signs of the eccentricity and elliptic-flow terms}\label{sec:origin}

The validation on states outside the fit motivates an examination of
why the freeze-out eccentricity $\efo$ enters Eq.~(\ref{eq:linear})
positively and the elliptic flow $v_2$ negatively. We first remove a velocity
gradient channel that cancels identically in the total harmonic. A
heavy-hadron saddle-point approximation then identifies the common response
of the remaining vorticity and shear contributions to the flow anisotropy.
Finally, an exact decomposition of the total integrand allows its
emission-gradient and curl terms to be compared with $\efo$ and $v_2$ on
the computed PICR surfaces. This sequence connects the empirical signs to
the velocity gradients and geometry at emission.

\subsection{Symmetry and locality}\label{sec:sym}

To identify which form of relation is compatible with the symmetry, consider
a pure second-harmonic perturbation of an axisymmetric background. A rotation
by $\pi/2$ reverses its amplitude and the signs of $\ein$, $\efo$, $v_2$,
and $\Pz$. Each observable is therefore odd under reversal of the perturbation; at
first order it is linear in the perturbation amplitudes and has no constant term. A relation involving only
$\efo$ and $v_2$ additionally requires the response of $\Pz$ to lie in the
span of their responses. The ILE prescription is local on $\Sigma$, so the
initial geometry influences $\Pz$ through the decoupling surface, the
velocity field and its gradients at emission, and the emission weights.
Symmetry and locality thus motivate a relation in freeze-out quantities;
the 48-state fit and the 12-state test establish how accurately $\efo$ and
$v_2$ describe the total harmonic in this calculation.

\subsection{Cancellation of the laboratory time-derivative channel (exact)}\label{sec:acc}

\begin{figure*}[t]
\centering
\includegraphics[width=0.95\textwidth]{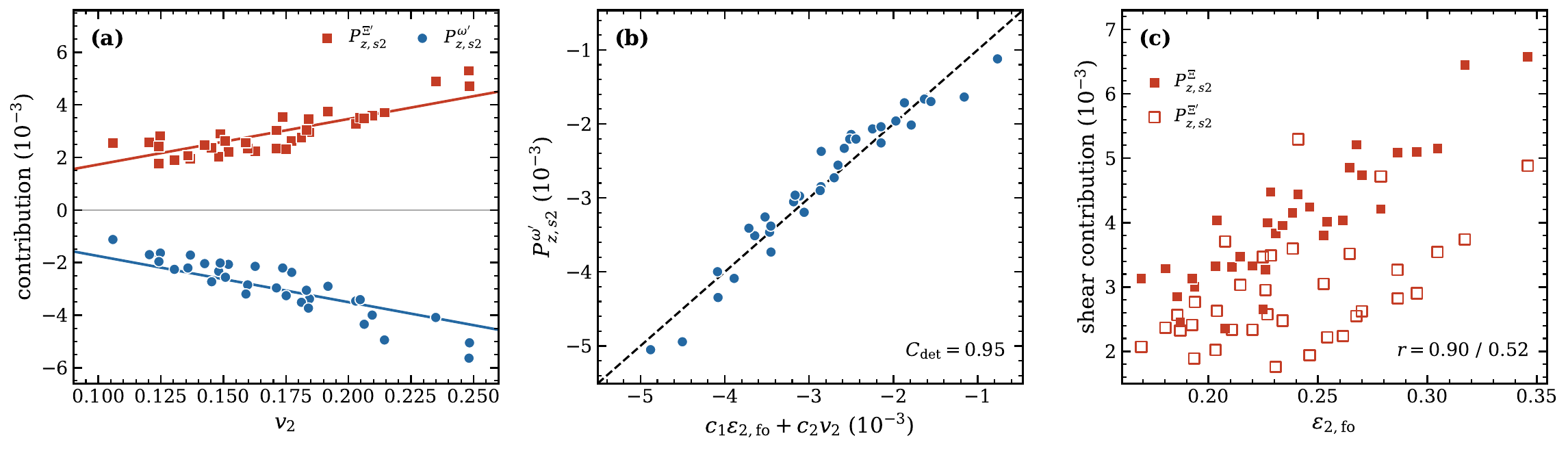}
\caption{Polarization contributions without the laboratory time-derivative
channel $\partial_tu_i$ for the 35 states with archived decoupling surfaces.
(a) $\Pz^{\Xi'}$ (squares) and $\Pz^{\omega'}$ (circles) versus the elliptic
flow $v_2$; lines are the one-variable fits of Table~\ref{tab:acc}. (b)
$\Pz^{\omega'}$ versus its two-variable fit $c_1\efo+c_2v_2$ with the coefficients of Table~\ref{tab:acc}; the dashed
line is equality. (c) Kinematic-shear contribution with ($\Pz^\Xi$, filled) and
without ($\Pz^{\Xi'}$, open) the $\partial_tu_i$ channel versus the freeze-out
eccentricity $\efo$.}
\label{fig:acc}
\end{figure*}

Equations~(\ref{eq:shear}) and (\ref{eq:vort}) assign the $\efo$ dependence to
$\Pz^\Xi$ and the $v_2$ dependence to $\Pz^\omega$. Before this assignment can
be interpreted, it must be checked against a regrouping of the velocity
gradients between the two terms of Eq.~(\ref{eq:ile}). The relevant gradients
are those with one time index. In the $z$ component of Eq.~(\ref{eq:ile}) they
enter the kinematic-vorticity term as $-\epsilon^{z0i\beta}p_\beta\,\omega_{0i}$
and the kinematic-shear term (the $\lambda=0$ part of
$\Xi_{\alpha\lambda}p^\lambda/E$) as $+\epsilon^{z0i\beta}p_\beta\,\Xi_{0i}$.
Their sum contains only
\begin{equation}
\Xi_{0i}-\omega_{0i}=\partial_iu_0,\qquad\text{whereas}\qquad
\Xi_{0i}+\omega_{0i}=\partial_0u_i .
\end{equation}
The laboratory-frame time derivatives $\partial_tu_i$ therefore enter
$\Pz^\omega$ and $\Pz^\Xi$ with opposite signs and cancel identically in
$\Pz$, as found analytically for anisotropic Gubser
flow~\cite{fukushima2026}. We denote by $\Pz^{(t)}$ the contribution of the
$\partial_tu_i$ channel to $\Pz^\Xi$ and define the two contributions without
this channel,
\begin{equation}
\Pz^{\omega'}=\Pz^\omega+\Pz^{(t)},\qquad
\Pz^{\Xi'}=\Pz^\Xi-\Pz^{(t)},
\end{equation}
whose sum is again $\Pz$. Because the integrand is linear in $\partial_\alpha u_\beta$, each gradient
channel can be integrated separately. Integrating the channels separately
requires the stored decoupling surfaces, which were archived for 35 of the 48 states (Appendix~\ref{app:channels}).
Table~\ref{tab:acc} lists one- and two-variable fits of each contribution over
these 35 states, and Fig.~\ref{fig:acc} shows the contributions without the
$\partial_tu_i$ channel.

\begin{table}[t]
\caption{Fits without intercept over the 35 states with archived decoupling
surfaces: one-variable fits $c\,v_2$ and two-variable fits
$c_1\efo+c_2v_2$, with signed coefficients in units of $10^{-3}$.}
\label{tab:acc}
\centering\footnotesize
\begin{tabular}{lcccc}
\toprule
contribution & $c$ & $\Cdet$ & $c_1,\ c_2$ & $\Cdet$\\
\midrule
$\Pz^\omega$ & $-23.0$ & 0.12 & $-9.8,\ -9.6$ & 0.66\\
$\Pz^\Xi$ & $+22.8$ & $<0$ & $20.3,\ -4.9$ & 0.80\\
$\Pz^{(t)}$ & $+5.5$ & $<0$ & $18.7,\ -19.9$ & 0.75\\
$\Pz^{\omega'}$ & $-17.5$ & 0.75 & $8.8,\ -29.6$ & 0.95\\
$\Pz^{\Xi'}$ & $+17.3$ & 0.72 & $1.6,\ 15.1$ & 0.73\\
$\Pz$ & $-0.2$ & 0.00 & $10.5,\ -14.5$ & 0.75\\
\bottomrule
\end{tabular}
\end{table}

Table~\ref{tab:acc} shows that most of the $\efo$ dependence of $\Pz^\Xi$
comes from the $\partial_tu_i$ channel: of the coefficient $20.3\times10^{-3}$,
$18.7\times10^{-3}$ belongs to $\Pz^{(t)}$, which is absent from $\Pz$.
Figure~\ref{fig:acc}(c) shows the same effect directly: the correlation of the
kinematic-shear contribution with $\efo$ drops from $0.90$ for $\Pz^\Xi$ to
$0.52$ for $\Pz^{\Xi'}$.

To see what remains after removing the $\partial_tu_i$ channel, we compare
the magnitudes of $\Pz^{\omega'}$, $\Pz^{\Xi'}$, and their sum on the 35
archived surfaces. $\Pz^{\omega'}$ and $\Pz^{\Xi'}$ are large and nearly cancel: on the
35 states $\Pz^{\omega'}$ ranges from $-5.6\times10^{-3}$ to $-1.1\times10^{-3}$ and
$\Pz^{\Xi'}$ from $+1.8\times10^{-3}$ to $+5.3\times10^{-3}$, whereas $|\Pz|$ is on average
only 9\% of $|\Pz^{\omega'}|+|\Pz^{\Xi'}|$.

To determine how $\Pz^{\omega'}$ and $\Pz^{\Xi'}$ vary with elliptic flow across the scan, we compare their one-variable fits in Fig.~\ref{fig:acc}(a) and Table~\ref{tab:acc}. The one-variable fits give
$\Pz^{\omega'}\simeq-17.5\times10^{-3}\,v_2$ and
$\Pz^{\Xi'}\simeq+17.3\times10^{-3}\,v_2$ [Fig.~\ref{fig:acc}(a)], with nearly
equal and opposite slopes. The slopes include the accompanying variation of $\efo$ across the sample. The fitted dependence on $v_2$ at fixed $\efo$ must therefore be determined
with both variables in the fit. Because these leading projections
on $v_2$ nearly cancel, $\Pz$ is determined by the variation that they do not
describe.

To identify the variation left after these opposite projections cancel,
we include the freeze-out eccentricity in the fits. They give
$\Pz^{\omega'}\simeq(8.8\,\efo-29.6\,v_2)\times10^{-3}$, with $\Cdet=0.95$
[Fig.~\ref{fig:acc}(b)], and $\Pz^{\Xi'}\simeq(1.6\,\efo+15.1\,v_2)\times10^{-3}$.
The two kinds of fit are consistent: with
$\kappa=\sum_iv_{2,i}\epsilon_{2,\mathrm{fo},i}/\sum_iv_{2,i}^2\simeq1.36$ on
these states, each one-variable slope equals the $v_2$ coefficient plus
$\kappa$ times the $\efo$ coefficient, so that
\begin{align}
\Pz^{\omega'}&\simeq\big[-17.5\,v_2+8.8\,(\efo-\kappa v_2)\big]\times10^{-3},\nonumber\\
\Pz^{\Xi'}&\simeq\big[+17.3\,v_2+1.6\,(\efo-\kappa v_2)\big]\times10^{-3}.
\end{align}
Their sum, up to rounding of the tabulated coefficients, is $\Pz\simeq[-0.2\,v_2+10.5\,(\efo-\kappa v_2)]\times10^{-3}
=(10.5\,\efo-14.5\,v_2)\times10^{-3}$, as in the last row of Table~\ref{tab:acc}.

The large projections on $v_2$ thus cancel, and $\Pz$ follows the deviation
$\efo-\kappa v_2$ of the freeze-out eccentricity from its average proportion to
$v_2$ in the sample; the near equality of $\kappa$ and $b/a$ restates the
vanishing one-variable slope of $\Pz$. Within this fitted relation, at fixed $v_2$ a larger $\efo$ pushes
$\Pz$ toward positive values, and across states the sign of $\Pz$ is set by
whether $\efo/v_2$ exceeds $b/a\simeq1.4$. This $\efo$ dependence enters mainly
through $\Pz^{\omega'}$. The power laws~(\ref{eq:shear}) and (\ref{eq:vort})
depend on the adopted division of $\Pz$, whereas $\Pz$ itself, and empirical
relations for it such as Eqs.~(\ref{eq:total}) and (\ref{eq:linear}), are
unaffected by this regrouping.

The vorticity power law~(\ref{eq:vort}) requires the survival fraction
$S_\epsilon$ in addition to $v_2$. Because an ideal-fluid estimate relates the
vorticity polarization to the cooling rate times $v_2$~\cite{becattini2018}, we
compare $S_\epsilon$ with the cooling rate $\lambda=-\langle\partial_t\ln T\rangle$,
the negative laboratory-time derivative of $\ln T$ at the emission point,
averaged with the
$\Lambda$ yield weight of Eq.~(\ref{eq:harm}). On the 35 decoupling surfaces,
$\lambda=0.11$--$0.14\,\mathrm{fm}^{-1}$, and its correlation coefficient with
$S_\epsilon$ is $0.88$, so that $S_\epsilon$ largely tracks the cooling rate. The
estimate based on vanishing T-vorticity itself does not reproduce the computed gradients
quantitatively (Appendix~\ref{app:channels}); the mechanism analysis below
therefore uses the computed gradients directly.

\subsection{Heavy-hadron saddle point (approximate)}\label{sec:saddle}

The projections of $\Pz^{\omega'}$ and $\Pz^{\Xi'}$ on $v_2$ are large and
nearly opposite. To identify the velocity gradient that produces these opposite
responses to the flow anisotropy, we evaluate
the two integrands near the emission point of a heavy hadron. For the transverse
integrand ($p_z\simeq0$) we use $\gamma=u^0$, the local-rest-frame energy $\chi=p\cdot u$, the longitudinal
component of the spatial curl $\Omega_z=\partial_xu^y-\partial_yu^x$, and the symmetric transverse four-velocity
gradient $\sigma_{ij}=\tfrac12(\partial_iu^j+\partial_ju^i)$ with
$i,j\in\{x,y\}$. The integrands of $\Pz^{\omega'}$ and $\Pz^{\Xi'}$ are then
\begin{align}
K^{\omega'}&=\tfrac12(\bm p\times\nabla\gamma)_z+\tfrac12E\,\Omega_z,\\
K^{\Xi'}&=\tfrac12(\bm p\times\nabla\gamma)_z-[\bm p\times(\sigma\bm p)]_z/E,
\end{align}
with $P^*_z=\langle K\rangle/(2mT)$. For a heavy hadron ($m/T\simeq7$ for the $\Lambda$ at $T=160\,\mathrm{MeV}$), the
emission at fixed momentum is concentrated where $\chi$ is small. If the surface
contains a flow-momentum matching point, $\bm u\simeq\bm p/m$, that dominates the
emission, we can use the saddle-point approximation of
Ref.~\cite{arslan2024}. A necessary condition is that $p_T/m$ lies within the
range of transverse four-velocities on $\Sigma$; at higher $p_T$, or when the
emission is dominated by the edge of the surface, the approximation does not
apply. At a matching point $\nabla\chi=0$, so that
$\nabla\gamma=(\nabla\bm u)\cdot\bm p/E$. We introduce streamline coordinates with unit vectors $\bm e_s$ along the transverse flow and $\bm e_n=\hat{\bm z}\times\bm e_s$ across the flow. Writing $\partial_n=\bm e_n\cdot\nabla$ and $q=|\bm u|$ for the magnitude of the spatial part of the four-velocity, with $p=|\bm p|$, gives
$[\bm p\times\sigma\bm p]_z=p^2(\partial_nq+\Omega_z/2)$, and therefore
\begin{equation}
K^{\omega'}\simeq\frac{p^2\partial_nq}{2E}+\frac{E\Omega_z}{2},\quad
K^{\Xi'}\simeq-\frac{p^2\partial_nq}{2E}-\frac{p^2\Omega_z}{2E}.
\label{eq:saddle}
\end{equation}

Equation~(\ref{eq:saddle}) isolates a common term proportional to the gradient $\partial_nq$ of the spatial four-velocity magnitude $q$ across streamlines. This term enters $K^{\omega'}$ and $K^{\Xi'}$ with opposite signs and cancels in the total integrand. For a nearly radial flow that is stronger in
the reaction plane, $q\propto1+2\delta\cos2\phi$ with $\delta>0$ along the
direction $\phi$ of the flow, one has
$\partial_nq\propto-\delta\sin2\phi$. The common $\partial_nq$ term therefore contributes negatively to $\Pz^{\omega'}$ and positively to $\Pz^{\Xi'}$, in proportion to the flow anisotropy $\delta$. A purely
geometric deformation does not produce this term if the flow is normal to the
elliptic contours and its magnitude $q$ is constant along each contour, because then
$\partial_nq=0$.

To check this identification of the common gradient on the PICR surfaces, we compare its
prediction for the $\nabla\gamma$ channel, built from the computed strain-rate and
curl channels, with the $\nabla\gamma$ channel evaluated directly from the stored gradients. The comparison gives a correlation coefficient of $0.92$ and a slope of $0.82$. Although
the slope shows that the local saddle-point approximation does not reproduce the
full amplitude, the strong correlation supports the gradient $\partial_nq$ of the spatial four-velocity magnitude across
streamlines as the source of the opposite flow responses of
$\Pz^{\omega'}$ and $\Pz^{\Xi'}$.

\subsection{Emission-gradient term and curvature of the decoupling surface}\label{sec:curv}

The saddle-point approximation explains why the common $\partial_nq$ term
cancels, but not which part of the remainder carries the $\efo$ dependence. To
answer this question we use an exact form of the total integrand. Summing all
gradient channels of Eq.~(\ref{eq:ile}) for on-shell momenta and arbitrary
velocity gradients gives
\begin{equation}
K=\frac{(\bm p\times\nabla\chi)_z}{E}+\frac{m}{2E}\Big[m\,\Omega_z+
\frac{p_z\,\bm p\cdot\bm\Omega}{E+m}\Big],
\label{eq:split}
\end{equation}
with spatial derivatives at fixed laboratory time and
$\bm\Omega=\nabla\times\bm u$, the spatial curl of the spatial components of the four-velocity (Appendix~\ref{app:identities}). The total integrand
thus separates into an emission-gradient term, which contains the gradient of
the local-rest-frame energy $\chi=p\cdot u$ that enters the Boltzmann weight $e^{-\chi/T}$, and a curl term, whose weight $m/2E$
depends on the hadron mass. On the PICR
surfaces these two terms reproduce $\Pz$ to $0.02\times10^{-3}$. The evaluation uses the computed velocity gradients and retains the $p_z$-dependent curl terms in Eq.~(\ref{eq:split}).

To determine how the emission-gradient and curl contributions depend on
$\efo$ and $v_2$, we first resolve the gradient of $\chi$ along and normal to
the decoupling surface. With $\hat n_\mu=d\Sigma_\mu/|d\Sigma|_E$ the Euclidean
unit normal in $(t,x,y,z)$ and $dA=|d\Sigma|_E$,
$\partial_\mu\chi=\nabla^\Sigma_\mu\chi+\hat n_\mu(\hat n\cdot\partial\chi)$.
The tangential projection and the surface curvatures used below are defined
with this Euclidean normal. The normal $\hat n$ refers to the decoupling surface, whereas $\bm e_n$ in
Sec.~\ref{sec:saddle} is the transverse direction across a streamline.
Table~\ref{tab:curv} compares the tangential emission gradient, the normal
emission gradient, and the curl contribution on the 35 archived surfaces.

The tangential contribution has a positive mean,
$+0.87\times10^{-3}$, and correlates most strongly with the freeze-out
eccentricity, $r(\efo)=0.82$. The normal emission gradient and curl term have
negative means; their sum has a mean of $-0.84\times10^{-3}$ and correlates
most strongly with the elliptic flow, $r(v_2)=-0.81$. The near cancellation
of these positive and negative contributions makes the total harmonic
sensitive to changes in their relative magnitudes.

To compare the geometric and flow dependence of each contribution, we also
fit both anisotropies simultaneously. The tangential contribution follows
$(5.1\efo-2.0v_2)\times10^{-3}$, while the normal emission gradient plus the
curl term follows $(5.4\efo-12.6v_2)\times10^{-3}$. Because $\efo$ and $v_2$
are only weakly correlated in this scan (correlation coefficient $0.33$), the two-variable fits can distinguish their associations with each
contribution. The tangential emission gradient and the sum of the normal
emission gradient and curl term contribute comparably to the geometric
coefficient $a$. Most of the flow coefficient $b$ comes from the normal
emission gradient and curl term. Refitting the tangential derivatives from the surface flow retains
the geometric association, with $r(\efo)=0.85$ and a median change of 3\%
of $|\Pz^\omega|+|\Pz^\Xi|$ (Appendix~\ref{app:channels}).

\begin{table}[t]
\caption{Exact split of the total harmonic $\Pz$ on the 35 PICR states into the
tangential and normal parts of the emission-gradient term and the curl term of
Eq.~(\ref{eq:split}). Means and signed fit coefficients of $c_1\efo+c_2v_2$ are
in units of $10^{-3}$; $r$ denotes correlation coefficients. $^{a}$Approximate
evaluation of the right-hand side of Eq.~(\ref{eq:curvature}) without the
boundary term, with the shape operator from local fits over $k=256$ facets
(Table~\ref{tab:knn}).}
\label{tab:curv}
\centering\footnotesize
\setlength{\tabcolsep}{3pt}
\begin{tabular}{lcccc}
\toprule
term & mean & $r(\efo)$ & $r(v_2)$ & $c_1,\ c_2$\\
\midrule
tangential emission gradient & $+0.87$ & $+0.82$ & $+0.27$ & $5.1,\ -2.0$\\
normal emission gradient & $-0.36$ & $+0.32$ & $-0.74$ & $3.8,\ -7.4$\\
curl term of Eq.~(\ref{eq:split}) & $-0.48$ & $+0.04$ & $-0.80$ & $1.7,\ -5.2$\\
normal $+$ curl & $-0.84$ & $+0.21$ & $-0.81$ & $5.4,\ -12.6$\\
\midrule
curvature form$^{a}$ & $+1.04$ & $+0.82$ & $+0.16$ & $7.4,\ -4.0$\\
\bottomrule
\end{tabular}
\end{table}

The strong correlation of the tangential contribution with $\efo$ motivates an explicit connection to the surface geometry. The emission-gradient term vanishes at a flow-momentum matching point, but its surface integral samples the finite emission region around that point. The shape of the decoupling surface determines how this region is weighted. To
make this dependence explicit, let $\ell=(0,-p_y,p_x,0)$ be a constant vector,
$S_{\mu\nu}=\nabla^\Sigma_\mu\hat n_\nu$ the shape operator of the surface, and
$\kappa_\Sigma$ its trace, the sum of the three principal curvatures. With constant $T$ on the isotherm, integration by
parts over $\Sigma$ gives for the tangential part
\begin{align}
&\int_\Sigma\! dA\,(p\cdot\hat n)\,f\,\frac{\ell\cdot\nabla^\Sigma\chi}{E}\nonumber\\
&=\frac{T}{E}\int_\Sigma\! dA\,f\Big[\ell^\mu S_{\mu\nu}p^\nu-\kappa_\Sigma\,(p\cdot\hat n)(\ell\cdot\hat n)\Big]\nonumber\\
&\quad-\frac{T}{E}\oint_{\partial\Sigma}\! dS\,f\,(p\cdot\hat n)(\ell\cdot\nu),
\label{eq:curvature}
\end{align}
where $dA\,(p\cdot\hat n)=p\cdot d\Sigma$, $dS$ is the two-dimensional area
element of the boundary $\partial\Sigma$, and $\nu$ is its outward co-normal,
tangent to $\Sigma$. Equation~(\ref{eq:curvature}) expresses the tangential
emission gradient through the curvature of the decoupling surface plus a
boundary term. The PICR surfaces have a boundary: they start at the initial time
slice, where the initial $T=T_{\mathrm{dec}}$ contour forms an edge, and a few
cells remain above $T_{\mathrm{dec}}$ at the last time step. The boundary term
vanishes if the surface meets the initial time slice perpendicularly, so that
$\ell\cdot\nu=0$. On the PICR surfaces the co-normal at the initial edge points
nearly along the time axis, and an approximate evaluation
(Appendix~\ref{app:channels}) gives a boundary term of $0.03\times10^{-3}$ on
average, 3--4\% of the mean tangential emission gradient; the edge at the last
time step contributes less than $10^{-6}$.

To test whether the tangential part indeed reflects the curvature of the
decoupling surface, we evaluate the right-hand side of Eq.~(\ref{eq:curvature})
approximately on the PICR surfaces, without the small boundary term and with the
shape operator of each facet obtained from a local fit of the facet normals over
its $k$ nearest facets (Appendix~\ref{app:channels}). For $k=256$ the curvature
form is positive in all 35 states, correlates with the tangential emission
gradient with $r=0.84$, and correlates with $\efo$ ($r=0.82$) but hardly with
$v_2$ ($r=0.16$), as the tangential part does (Table~\ref{tab:curv}); its mean is
1.2 times the tangential mean. Increasing the neighborhood size reduces the effect of noisy Cornelius facet normals, and the mean decreases with $k$ (Table~\ref{tab:knn}): on 12 of the 35 states with
archived surfaces, $k=512$ brings the mean to within 5\% of the tangential mean,
or 10\% with the boundary term added, and raises the correlation to $0.91$. Although the discrete estimate has not fully converged, the recovered sign and geometric correlation, together with the agreement of the means on the 12-state subset, strongly support the curvature origin of the positive tangential contribution.

Taken together, the exact split~(\ref{eq:split}), the saddle-point
relation~(\ref{eq:saddle}), and Table~\ref{tab:curv} give the following picture
of the signs in Eq.~(\ref{eq:linear}). When heavy-hadron emission is dominated by flow-momentum matching, the same gradient $\partial_nq$ of the spatial four-velocity magnitude $q$ across streamlines produces opposite contributions to $\Pz^{\Xi'}$ and $\Pz^{\omega'}$ and cancels from their sum. Because the
emission region has a finite width, the remaining emission gradient samples
the shape of the decoupling surface, and Eq.~(\ref{eq:curvature}) identifies
how the surface geometry enters its tangential part. The PICR calculation
shows that this tangential part is positive on average and most strongly
correlated with $\efo$, and the direct evaluation of the curvature form
reproduces its sign, its $\efo$ dependence, and, on the 12-state subset with $k=512$, its mean to about 10\% including the estimated boundary term. The
opposing negative contribution comes from the normal emission gradient and
from the spatial-curl contribution, and is most strongly correlated with
$v_2$.

\section{Dependence of the coefficients on the decoupling background in a blast-wave model}\label{sec:bw}

Section~\ref{sec:origin} interprets the signs of the $\efo$ and $v_2$ terms but not the magnitudes of $a$ and $b$ or the cancellation point $b/a\simeq1.4$. To
test whether the structure of the integrand found in Sec.~\ref{sec:origin} also produces these
magnitudes, we evaluate the full integrand on a boost-invariant blast-wave
background in which the curvature of the decoupling surface and the flow can be
varied independently. The background is kinematic: we prescribe the
temperature $T(\tau,x,y)$ and the transverse four-velocity
$\bm U(\tau,x,y)$ in a neighborhood of the surface and take all derivatives
from these fields (Appendix~\ref{app:bw}).

The temperature is
$\ln T=\ln T_f+\beta[1-(\tau-\tau_0)^2/\Delta\tau^2-x^2/r_x^2-y^2/r_y^2]$.
The decoupling surface $T=T_f$ is then the half-ellipsoid
$\tau_s(x,y)=\tau_0+\Delta\tau\sqrt{1-x^2/r_x^2-y^2/r_y^2}$, independent of
$\eta_s$, with $\tau_0=4\,\mathrm{fm}/c$; the surface is space-like at the center (time-like normal) and
time-like near the rim (space-like normal). The time depth $\Delta\tau$ is the proper-time interval by
which the center decouples after the rim. The coefficient $\beta=|\lambda_c|\Delta\tau/2$ is fixed by
the signed logarithmic temperature derivative at the center,
$\lambda_c=\partial_\tau\ln T=-(1/\tau_c+2H)/3<0$, where $1/\tau_c$ is a
separate parameter for the longitudinal expansion rate.

The transverse flow is derived from a time-independent potential,
$\bm U=T_f\nabla\Phi/T$ with $\Phi=(H_xx^2+H_yy^2)/2$, so that
$\bm U=(H_xx,H_yy)$ on the surface. The derivatives are
$\partial_iU^j=(T_f/T)\partial_i\partial_j\Phi-U^j\partial_i\ln T$ at fixed
$\tau$ and $\partial_\tau U^j=-U^j\partial_\tau\ln T$. Because
$\partial_i\ln T+\partial_\tau\ln T\,\partial_i\tau_s=0$ on the isotherm, these
derivatives reproduce the derivative of the surface flow along the surface. The
longitudinal flow is the Bjorken flow. The spatial and flow anisotropies enter
through $r_{x,y}=r_0\sqrt{1\mp\epsilon_R}$ and $H_{x,y}=H(1\pm\delta)$.

To determine the polarization coefficients for this background, we evaluate the full laboratory-frame integrand, Eq.~(\ref{eq:split}) including the
$p_z$ terms, with the signed Cooper-Frye weight, compute $\efo$, $v_2$, and
$\Pz$ with the definitions of Sec.~\ref{sec:model}, and eliminate $(\epsilon_R,\delta)$ at first order. The first-order responses satisfy
\begin{align}
\begin{pmatrix}\efo\\ v_2\end{pmatrix}&=M\begin{pmatrix}\epsilon_R\\ \delta\end{pmatrix},\qquad
\Pz=J\begin{pmatrix}\epsilon_R\\ \delta\end{pmatrix},\nonumber\\
(a,-b)&=J\,M^{-1},
\label{eq:bwresp}
\end{align}
where $M$ is the $2\times2$ response matrix of $(\efo,v_2)$ and $J$ is the row of responses of $\Pz$. The matrix $M$ is invertible, with condition number about 2; Appendix~\ref{app:bw} gives the numerical implementation. For an invertible response matrix, eliminating the two deformation
parameters fixes the linear form. The comparison with PICR therefore tests
the signs and magnitudes of $a$ and $b$, and their ratio, without using the
PICR polarization to calibrate the blast-wave background.

The model satisfies the transverse condition $\nabla_\perp\times(T\bm U)=0$ at
fixed $\tau$, which gives the curl
\begin{align}
\Omega_z&=(\bm U\times\nabla\ln T)_z
=-\frac{2\beta H}{r_0^2}\,xy\left[\frac{1+\delta}{1+\epsilon_R}-\frac{1-\delta}{1-\epsilon_R}\right]\nonumber\\
&\simeq-\frac{2\beta H}{r_0^2}\,r_s^2\sin2\phi_s\,(\delta-\epsilon_R),
\label{eq:bwcurl}
\end{align}
where $(r_s,\phi_s)$ are polar coordinates in the transverse plane and the
second line is first order in the anisotropies. Equation~(\ref{eq:bwcurl}) is
the analogue of the blast-wave result that the longitudinal vorticity is
proportional to the difference between the flow anisotropy $\delta$ and the
spatial anisotropy $\epsilon_R$~\cite{voloshin2017}. To assess the sensitivity to the transverse components $\Omega_x$ and $\Omega_y$ of the spatial curl, we replace them by the values implied by vanishing T-vorticity, $d(Tu)=0$, holding all other quantities fixed (Appendix~\ref{app:bw}). Under this algebraic substitution, $b/a$ changes by about $0.2$, while both $a$ and $b$ remain positive.

\begin{figure*}[t]
\centering
\includegraphics[width=0.95\textwidth]{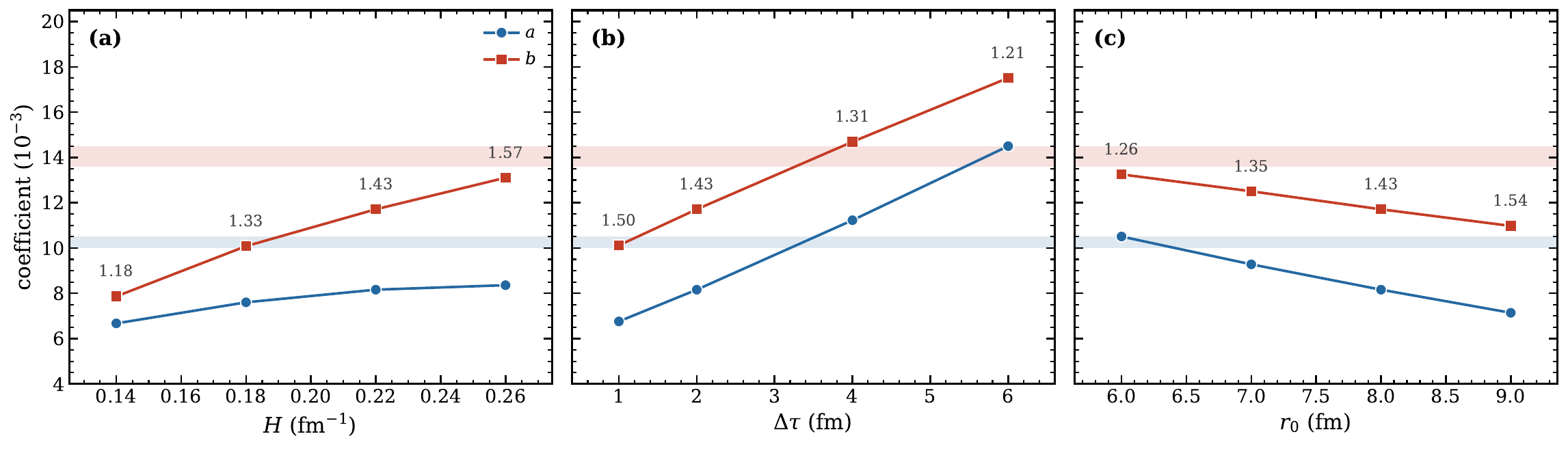}
\caption{Blast-wave coefficients $a$ (circles) and $b$ (squares) of
Eq.~(\ref{eq:linear}) versus the flow strength $H$, the time depth
$\Delta\tau$ of the decoupling surface, and the transverse size $r_0$, with the
other parameters at the reference values of Table~\ref{tab:bw}. Numbers above
the squares give $b/a$. The shaded bands mark the PICR fits of $a$ and $b$ on
the 48 and 35 states.}
\label{fig:bw}
\end{figure*}

To compare backgrounds with the same decoupling temperature and similar
expansion rates, we set $T_f=T_{\mathrm{dec}}=160\,\mathrm{MeV}$ and choose the
flow strength $H$ so that the expansion-rate estimate $\theta=1/\tau_c+2H$, which sets
the cooling of the background, matches the yield-weighted expansion rate on the
PICR decoupling surfaces,
$\theta\simeq0.5$--$0.65\,\mathrm{fm}^{-1}$; the transverse size $r_0$ and
the time depth $\Delta\tau$ are varied over ranges comparable to the PICR
surfaces. The reference point ($H=0.22\,\mathrm{fm}^{-1}$,
$\Delta\tau=2\,\mathrm{fm}$, $r_0=8\,\mathrm{fm}$, $\tau_c=10\,\mathrm{fm}$) is one
choice within these ranges. Table~\ref{tab:bw} and Fig.~\ref{fig:bw} show how
$a$, $b$, and $b/a$ depend on these parameters.

\begin{table}[t]
\caption{Blast-wave coefficients $a$ and $b$ (in units of $10^{-3}$) and their
ratio; numerical convergence is better than 3\%. $^{a}$Reference point:
$H=0.22\,\mathrm{fm}^{-1}$, $\Delta\tau=2\,\mathrm{fm}$, $r_0=8\,\mathrm{fm}$,
$\tau_c=10\,\mathrm{fm}$, $\tau_0=4\,\mathrm{fm}$, linear flow profile, and
$T_f=160\,\mathrm{MeV}$; the following rows change one parameter. The row
``varied'' covers $\tau_c=6$--$14\,\mathrm{fm}$, flow profiles
$U\propto r_s^{0.5}$--$r_s^{2}$, and $T_f=155$--$170\,\mathrm{MeV}$.
$^{b}$Algebraic sensitivity check: spatial-curl components $\Omega_x$ and $\Omega_y$ replaced by
the values implied by $d(Tu)=0$, all other quantities unchanged.}
\label{tab:bw}
\centering\footnotesize
\setlength{\tabcolsep}{3pt}
\begin{tabular}{lccc}
\toprule
parameters & $a$ & $b$ & $b/a$\\
\midrule
reference$^{a}$ & 8.2 & 11.7 & 1.43\\
$\Delta\tau=4\,\mathrm{fm}$ & 11.2 & 14.7 & 1.31\\
$H=0.18$ / $0.26\,\mathrm{fm}^{-1}$ & 7.6 / 8.4 & 10.1 / 13.1 & 1.33 / 1.57\\
$r_0=6$ / $9\,\mathrm{fm}$ & 10.5 / 7.1 & 13.3 / 11.0 & 1.26 / 1.54\\
$\tau_c$, profile, $T_f$ varied & 8.0--8.5 & 11.5--12.2 & 1.42--1.50\\
reference, curl check$^{b}$ & 8.6 & 10.5 & 1.22\\
PICR fit (35 states) & 10.5 & 14.5 & 1.38\\
\bottomrule
\end{tabular}
\end{table}

Table~\ref{tab:bw} and Fig.~\ref{fig:bw} allow the coefficient magnitudes
to be compared at fixed values of all but one background parameter. For
$H=0.18$--$0.26\,\mathrm{fm}^{-1}$, the expansion-rate estimate is
$\theta=0.46$--$0.62\,\mathrm{fm}^{-1}$, overlapping the PICR range, and
$b/a=1.33$--$1.57$ brackets the PICR value 1.38. At the reference point,
$a=8.2\times10^{-3}$ and $b=11.7\times10^{-3}$ have the same signs and
comparable magnitudes as the PICR coefficients; increasing the time depth to
$\Delta\tau=4\,\mathrm{fm}$ brings both within 7\% of the 35-state fit.
A calculation restricted to the space-time-rapidity slice $\eta_s=0$ gives
$b/a\simeq0.75$--$0.86$ over the flow profiles examined, showing the
importance of the space-time-rapidity integral for $b$. Within the ranges
examined, $a$ and $b$ respond mainly to the flow strength $H$, time depth
$\Delta\tau$, and transverse size $r_0$; changes of $\tau_c$, $T_f$, and the
radial profile alter them by less than 5\%. Thus, a background specified
without polarization input can reproduce the coefficient scale, while the
parameter dependence identifies which freeze-out properties shift the
cancellation point.

The agreement at the reference parameters motivates a stricter comparison:
we determine the blast-wave parameters separately from each PICR surface,
without using its polarization (Appendix~\ref{app:bw}). Two prescriptions for the source anisotropy, taking the source radii from the spatial second moments of the surface or choosing $\epsilon_R$ to reproduce the PICR $\efo$, give $b/a\simeq0.6$--$0.9$. With the first prescription, the mapped elliptic flow and polarization retain the PICR ordering, with correlation coefficients of $0.99$ and $0.83$, respectively. The mapped surfaces have weaker flow, a different
time depth, and a different transverse size from the reference background,
all of which affect $b/a$ in Table~\ref{tab:bw}. Repeating six states with an earlier start, $t_0=2\,\mathrm{fm}/c$, and the rescaled initial energy-density normalization specified in Appendix~\ref{app:bw} strengthens the flow
and changes the surface geometry: the mapped and PICR cancellation points
move from 0.63 and 1.34 to 0.88 and 1.05, respectively. Although the simple mapping still does not fully reproduce PICR, the
closer cancellation points support the importance of the freeze-out
background in setting the balance between the geometric and flow terms.

The integrand~(\ref{eq:split}) depends on the hadron mass, which suggests a test
with other baryons. Repeating the calculation at the reference point gives
$b/a=1.36$ ($m_p$), $1.43$ ($m_\Lambda$), $1.52$ ($m_\Xi$), and $1.67$
($m_\Omega$) (all with the spin-$\tfrac12$ integrand; for the spin-$\tfrac32$
$\Omega$, the factor $S(S+1)/3$ multiplies both mean-spin terms; after division by
$S$ to obtain the polarization, $a$ and $b$ increase by $5/3$, while $b/a$ is unchanged~\cite{becattini2017,palermo2024}). In this
spin-$\tfrac12$ mass scan, the common prefactor $1/(2mT)$ scales $a$
and $b$ alike, and $a\,m$ indeed varies by only 3\%. By contrast, $b\,m$ rises
by 26\%, because the curl contribution to $b$ stays nearly constant
($3.3$--$3.5\times10^{-3}$), whereas the emission-gradient contribution falls
roughly as $1/m$. At a common decoupling surface, heavier hyperons are
therefore expected to have a higher cancellation point $\efo/v_2=b/a$.

To test this mass trend on hydrodynamic surfaces, we repeat the ILE
calculation for the same masses on the 35 PICR states and fit
Eq.~(\ref{eq:linear}) using the elliptic flow of each hadron.
Table~\ref{tab:mass} shows that $b/a$ rises from 1.28 for the proton mass
to 1.63 for the $\Omega$ mass; the 68\% bootstrap intervals of neighboring
masses do not overlap. As in the blast-wave calculation, $a\,m$ varies by only
3\%, whereas $b\,m$ rises by 30\%. The $\Omega$-to-$\Lambda$ ratio of
cancellation points is 1.18 in PICR and 1.17 in the blast-wave background,
whose individual cancellation points are 0.04--0.08 higher. Although earlier
decoupling of multistrange hadrons could change the quantitative comparison,
the common-surface calculations in both backgrounds support a rising
cancellation point with mass. Polarization and elliptic-flow measurements for $\Xi$ and $\Omega$
can therefore test this mass dependence of the cancellation boundary.

\begin{table}[t]
\caption{Coefficients of Eq.~(\ref{eq:linear}) on the 35 PICR states for
hadrons of different mass, all evaluated with the spin-$\tfrac12$ integrand and
fitted with the elliptic flow of each hadron ($a$ and $b$ in units of
$10^{-3}$), with 68\% bootstrap intervals of $b/a$, compared with the blast-wave
reference point. For the spin-$\tfrac32$ $\Omega$, the polarization is the mean spin divided
by the particle spin; its $a$ and $b$ are larger by the factor $5/3$, and $b/a$ is unchanged. The momentum grid is
coarser than in Sec.~\ref{sec:empirical}; for the $\Lambda$ it reproduces
$a=10.5$ and $b=14.5$ within 2\%.}
\label{tab:mass}
\centering\footnotesize
\setlength{\tabcolsep}{2pt}
\begin{tabular}{lcccccc}
\toprule
 & $m$ (GeV) & $a$ & $b$ & $b/a$ & $\Cdet$ & $b/a$ (blast wave)\\
\midrule
$p$ & 0.938 & 12.8 & 16.4 & 1.28 (1.25--1.32) & 0.74 & 1.36\\
$\Lambda$ & 1.116 & 10.7 & 14.7 & 1.38 (1.35--1.41) & 0.75 & 1.43\\
$\Xi$ & 1.322 & 9.0 & 13.4 & 1.48 (1.45--1.52) & 0.76 & 1.52\\
$\Omega$ & 1.672 & 7.3 & 12.0 & 1.63 (1.59--1.68) & 0.77 & 1.67\\
\bottomrule
\end{tabular}
\end{table}

\section{Freeze-out geometry from ALICE polarization data}\label{sec:scope}

The coefficient variations in Sec.~\ref{sec:bw} determine how measured
polarization can constrain the freeze-out geometry. For a specified background,
Eq.~(\ref{eq:linear}) separates the eccentricity required to balance the flow
term from the additional eccentricity associated with a nonzero $\Pz$. The PICR coefficients are calibrated on ideal hydrodynamics with $J_y=0$
and primary $\Lambda$ hyperons. Viscosity~\cite{palermo2025}, longitudinal structure with
nonzero $J_y$~\cite{ryu2021,alzhrani2022}, and resonance
feed-down~\cite{bcs2019,xia2019} can change the mapping between measured
polarization and geometry. Although these effects are outside the present
calibration, comparing specified PICR and blast-wave backgrounds determines
what geometric information the ALICE data provide within the ILE prescription
and how that inference changes with the freeze-out conditions.

\subsection{Inversion and its dependence on the background}

If the coefficients of Eq.~(\ref{eq:total}) are taken from PICR, a measured
positive $\Pz$, a measured $v_2$, and an interval of $\ein$ determine $\efo$ on
the shear-dominated branch of Eq.~(\ref{eq:total}). We use the compact relation
for the main inversion because it contains the initial eccentricity $\ein$
explicitly, so that the inferred $\efo$ itself shows how the assumed initial
geometry enters; the linear relation~(\ref{eq:linear}) gives a cross-check of $\efo$ without
an initial-eccentricity input. Combining that linear-inversion value of
$\efo$ with the same interval of $\ein$ then gives the survival fraction
$S_\epsilon=\efo/\ein$.

We apply the compact relation to ALICE data; Table~\ref{tab:inputs} lists the inputs. The polarization
harmonics are the ALICE measurements in Pb-Pb collisions at
$\sqrt{s_{\mathrm{NN}}}=5.02$~\cite{alice2022} and
$5.36\,\mathrm{TeV}$~\cite{alice536} for $|y|<0.5$ and
$0.5<p_T<6\,\mathrm{GeV}/c$. The $5.36\,\mathrm{TeV}$ values were extracted
from the vector graphics of the published figure and calibrated with the
overlaid $5.02\,\mathrm{TeV}$ points. For both energies we use the ALICE
$\Lambda$ elliptic flow at $5.02\,\mathrm{TeV}$~\cite{alicev2}, weighted with
the measured $\Lambda$ spectrum~\cite{alicespectra} over
$0.6<p_T<6\,\mathrm{GeV}/c$, as the flow input, and vary it by $\pm5\%$ to test the sensitivity to using
this proxy at $5.36\,\mathrm{TeV}$. In the weighting, statistical uncertainties are added in
quadrature, and systematic uncertainties are treated as fully correlated in
$p_T$. The interval of $\ein$ is the range over the six PICR transverse
smoothing widths at the reduced impact parameter $b_0$ given by $c\simeq b_0^2$,
interpolated linearly between $b_0=0.60$ and $0.77$; for 30--40\% centrality,
$b_0$ is set to the lower end $0.60$.

\begin{table*}[t]
\caption{Inputs and results of the inversion of Eq.~(\ref{eq:total}). ALICE
polarization harmonic $\Pz$ in units of $10^{-3}$ with statistical and
systematic uncertainties; spectrum-weighted $\Lambda$ elliptic flow $v_2$ at
$5.02\,\mathrm{TeV}$ with total uncertainty (used as proxy at
$5.36\,\mathrm{TeV}$); interval of the initial eccentricity $\ein$; and the
resulting range of the freeze-out eccentricity $\efo$ for the central values of
the inputs. The
ranges including experimental uncertainties are given in the text.}
\label{tab:inputs}
\centering\footnotesize
\setlength{\tabcolsep}{6pt}
\begin{tabular}{llcccc}
\toprule
$\sqrt{s_{\mathrm{NN}}}$ (TeV) & centrality & $\Pz$ & $v_2$ & $\ein$ & $\efo$\\
\midrule
5.02 & 30--40\% & $0.80\pm0.15\pm0.15$ & $0.119\pm0.002$ & 0.33--0.46 & 0.245--0.270\\
5.02 & 40--50\% & $1.00\pm0.20\pm0.25$ & $0.129\pm0.001$ & 0.33--0.49 & 0.271--0.306\\
5.36 & 30--40\% & $0.46\pm0.08\pm0.05$ & (0.119) & 0.33--0.46 & 0.216--0.233\\
5.36 & 40--50\% & $0.98\pm0.10\pm0.10$ & (0.129) & 0.33--0.49 & 0.270--0.305\\
\bottomrule
\end{tabular}
\end{table*}

\begin{figure*}[t]
\centering
\includegraphics[width=0.86\textwidth]{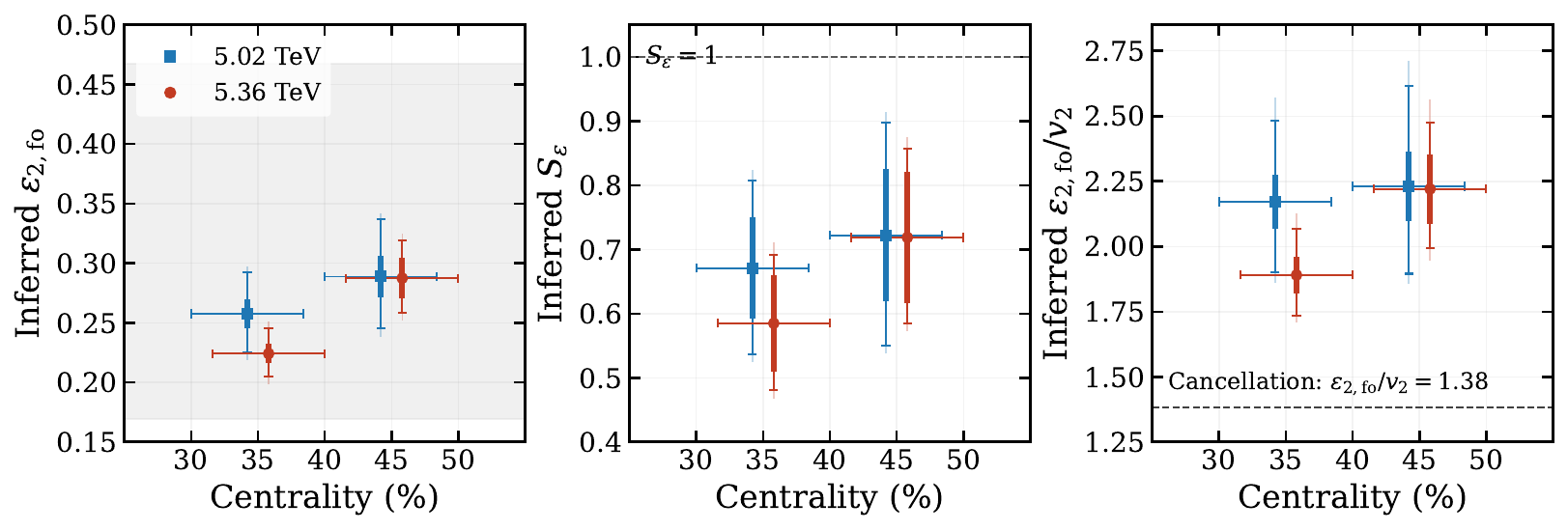}
\caption{Geometric quantities inferred by inverting Eq.~(\ref{eq:total}) for
ALICE longitudinal-polarization data in Pb-Pb collisions at 5.02 and
$5.36\,\mathrm{TeV}$: the freeze-out eccentricity $\efo$, the survival fraction
$S_\epsilon$, and $\efo/v_2$. Blue squares denote $5.02\,\mathrm{TeV}$ and red
circles $5.36\,\mathrm{TeV}$. Thick bars show the range over the interval of
the initial eccentricity $\ein$ for the central values of the inputs. Capped
bars additionally
include the experimental uncertainties of $\Pz$ and $v_2$. Faint bars further
include the $\pm5\%$ variation of the elliptic-flow proxy. The gray band is the
$\efo$ range of the 48-state scan; dashed lines mark $S_\epsilon=1$ and the
cancellation boundary $\efo/v_2=1.38$. All bars use the global PICR
coefficients and do not include the calibration scatter of the relation; the
change of $\efo$ with the background is discussed in the text.}
\label{fig:inversion}
\end{figure*}

Figure~\ref{fig:inversion} shows the geometry inferred with the global PICR
coefficients. At $5.36\,\mathrm{TeV}$, the central measurement values give
$\efo=0.216$--$0.233$ in 30--40\% and $0.270$--$0.305$ in 40--50\% collisions.
Including the experimental uncertainties widens these ranges to
$0.205$--$0.245$ and $0.258$--$0.320$; adding the $v_2$ proxy variation gives
$0.198$--$0.251$ and $0.252$--$0.325$. The linear relation gives
$\efo=(\Pz+b\,v_2)/a=0.21$ and $0.27$ for 30--40\% and 40--50\% collisions,
respectively, using the central inputs. These values are close to the
compact-relation ranges, lie within the $\efo$ range of the 48-state scan,
and give a survival fraction $S_\epsilon<1$.

The bars in Fig.~\ref{fig:inversion} propagate the input variations at fixed
global coefficients. The scatter of the empirical relation adds a separate
uncertainty in the inferred geometry: the root-mean-square error
$0.28\times10^{-3}$ of the linear relation on the eight additional states
corresponds to an eccentricity scale of about $0.03$ after division by $a$.
The propagation of coefficient uncertainties for the local calibrations is
given in the Supplemental Material~\cite{supp}.

To identify the geometric information added by polarization, we compare the
linear inversion with the cancellation boundary $\efo=(b/a)v_2$. The
measured $\Pz$ determines the displacement from this line,
\begin{equation}
\Delta\epsilon_{2,\mathrm{fo}}\equiv\efo-\frac ba\,v_2=\frac{\Pz}{a},
\label{eq:offset}
\end{equation}
which is positive for the ALICE values used here. With the global linear
coefficients, the displacement is $0.046$ in 30--40\% and $0.098$ in 40--50\%
collisions at $5.36\,\mathrm{TeV}$. Thus, once $v_2$ and the background
coefficients are specified, polarization determines how much additional
freeze-out eccentricity is needed beyond the value that cancels the flow term.

Section~\ref{sec:bw} shows that the background changes both the scale $a$ of the geometric displacement and the cancellation point $b/a$. To quantify the resulting variation of the inferred $\efo$ within PICR, we recalibrate
Eq.~(\ref{eq:linear}) locally, on six PICR states for each centrality class at
the reduced impact parameter matched by $c\simeq b_0^2$ ($b_0=0.60$ for 30--40\%
and $0.685$ for 40--50\%), with transverse smoothing widths chosen so that $v_2$
brackets the measured values, once with the original start time
$t_0=4\,\mathrm{fm}/c$ and once with $t_0=2\,\mathrm{fm}/c$, with the initial energy-density normalization $e_{\mathrm{peak}}$ rescaled according to Appendix~\ref{app:bw}.
At $5.36\,\mathrm{TeV}$ the $t_0=4$ and $2\,\mathrm{fm}/c$ calibrations give, respectively,
$\efo\simeq0.27$ and $0.19$ in 30--40\% collisions, and $0.33$ and $0.25$
in 40--50\% collisions. The earlier
start, with its stronger flow, thus lowers the inferred $\efo$ by about 25\%,
more than the experimental uncertainties.

To compare with a spectrum-constrained background, we adopt the radial flow
profile and kinetic freeze-out temperature
$T_{\mathrm{kin}}\simeq0.10$--$0.11\,\mathrm{GeV}$ from the ALICE blast-wave
fit to $\pi^\pm$, $K^\pm$, and (anti)proton spectra in Pb-Pb collisions at
$5.02\,\mathrm{TeV}$~\cite{alicebw}. We compute the ILE polarization at
$T_f=T_{\mathrm{kin}}$, whereas PICR uses $T_{\mathrm{dec}}=160\,\mathrm{MeV}$,
close to the chemical freeze-out temperature~\cite{andronic2018}. Our
direct-$\Lambda$ mean transverse momentum in $0.6<p_T<6\,\mathrm{GeV}/c$
is 5--10\% above the ALICE measurement. Varying the source radius and time
depth, which the spectra do not fix, gives $b/a=1.3$--$1.7$ and
$\efo=0.18$--$0.28$ in 30--40\% and $0.23$--$0.39$ in 40--50\% collisions
(Supplemental Material~\cite{supp}).

We compare the global and local PICR inversions and the spectrum-constrained
blast-wave inversions at fixed central ALICE inputs at $5.36\,\mathrm{TeV}$.
Although the background choice changes the inferred $\efo$ by about 30\%,
the global and local PICR values fall within the blast-wave ranges,
$\efo=0.18$--$0.28$ in 30--40\% and $0.23$--$0.39$ in 40--50\% collisions.
The measured polarization therefore constrains freeze-out geometry within
the ILE prescription both for PICR emission at $160\,\mathrm{MeV}$ and for
the examined blast-wave backgrounds at
$T_f=T_{\mathrm{kin}}\simeq100$--$110\,\mathrm{MeV}$. Independent constraints
on the flow and decoupling surface can reduce the background dependence of
that inference.

\subsection{Comparison with femtoscopic eccentricities}

To compare the geometry inferred from polarization with an independent
observable, we consider azimuthally differential pion femtoscopy. Its estimator $\epsilon_{\mathrm{final}}\simeq2R^2_{s,2}/R^2_{s,0}$ uses
the mean $R^2_{s,0}$ and second Fourier coefficient $R^2_{s,2}$ of the
squared side radius, defined by
$R_s^2(\Phi)=R^2_{s,0}+2R^2_{s,2}\cos2\Phi+\cdots$, with pair azimuth
$\Phi$ measured from the reaction plane; at small pair momentum it
estimates the source eccentricity to about
20--30\%~\cite{retiere2004}. ALICE measurements in Pb-Pb collisions at
$2.76\,\mathrm{TeV}$ give $\epsilon_{\mathrm{final}}\approx0.05$--$0.12$
in 30--40\% and $0.08$--$0.16$ in 40--50\% centrality, retaining about
0.2--0.4 of the Glauber initial eccentricity~\cite{alicehbt}. These ranges,
read from Figs.~3 and 4 of Ref.~\cite{alicehbt}, are below the polarization
inversion ranges obtained from the central measurement values above.

The comparison involves different collision energies and different emission
weights: pion femtoscopy selects a momentum-dependent homogeneity region,
whereas $\efo$ weights the whole surface by the fluid flux $u^\mu d\Sigma_\mu$. We therefore
evaluate the pion estimator on the same PICR surfaces as the polarization
(Supplemental Material~\cite{supp}). For direct thermal pions at pair transverse momentum
$k_T=0.25\,\mathrm{GeV}/c$ it averages 0.91 of $\efo$, retains 0.39--0.63
of $\ein$, and describes $\Pz$ less accurately than $\efo$ when substituted
in the fitted relation. The same-surface comparison therefore shows that the fluid-flux-weighted eccentricity $\efo$ and the pion femtoscopic estimator retain different geometric information even for one hydrodynamic source. A joint calculation at matched centralities and collision energies could use polarization and pion femtoscopy to constrain both the source geometry and its emission profile.

\section{Summary}\label{sec:summary}

For the ideal-hydrodynamic Pb-Pb scan and ILE prescription studied in this
paper, the competition between the negative kinematic-vorticity contribution
$\Pz^\omega$ and positive kinematic-shear contribution $\Pz^\Xi$ is
quantified by $\Pz\simeq a\efo-bv_2$. In the fitted relation, increasing the freeze-out eccentricity $\efo$ at fixed $v_2$ raises $\Pz$, whereas increasing the elliptic flow $v_2$ at fixed $\efo$ lowers $\Pz$. Calibrated on 48 states, the relation describes 86\% of
the variation in 12 states outside the fit. The
cancellation point $\efo/v_2\simeq1.4$ is also reproduced by the compact
empirical form, linking the sign change to the balance of spatial and
momentum anisotropy at emission.

The physical interpretation combines identities of the ILE integrand with
the computed gradient responses. Laboratory time derivatives cancel exactly
between $\Pz^\omega$ and $\Pz^\Xi$, and the total integrand separates exactly
into an emission-gradient term and a curl term. Where flow-momentum
matching dominates heavy-hadron emission, the saddle-point approximation
identifies the gradient $\partial_nq$ of the spatial four-velocity magnitude $q$ across streamlines, which enters $\Pz^{\omega'}$ and $\Pz^{\Xi'}$ with opposite signs. On the PICR surfaces, the remaining
tangential emission gradient is positive on average and correlates most
strongly with $\efo$; the normal emission gradient and curl term are
negative on average and correlate most strongly with $v_2$. Although the
discrete curvature estimate has not fully converged, it reproduces the
tangential mean within about 10\%, including the estimated boundary term,
on 12 of the 35 states with archived surfaces. Together with the exact curvature identity, this
agreement strongly supports the geometric origin of the positive tangential
contribution. The two-variable relation is validated by the hydrodynamic
scan and additional states, while the decomposition explains the physical
competition represented by its coefficients.

Because the surface analysis explains the signs but not the magnitudes of the
coefficients, the blast-wave comparison examines how they depend on the
freeze-out background. Reference parameters chosen without polarization input give the same signs
and comparable coefficient magnitudes as PICR, while the state-by-state
mapping yields a lower cancellation point. The parameter scan and mapping
identify the flow strength, time depth, and transverse size as freeze-out
properties that shift the balance between the geometric and flow terms. On common decoupling surfaces, both
models give a cancellation point that rises with mass, from about 1.3 for
the proton to 1.6 for the $\Omega$ in PICR (the spin-$\tfrac32$ factor of the
$\Omega$ cancels in $b/a$). This provides a specific trend
for $\Xi$ and $\Omega$ polarization studies with their decoupling conditions
taken into account.

Although the inferred geometry depends on the freeze-out background,
inverting the central ALICE measurement values at $5.36\,\mathrm{TeV}$ gives
$\efo$ of about 0.2--0.4 in 30--50\% Pb-Pb collisions for PICR emission at
$160\,\mathrm{MeV}$ and for the spectrum-constrained blast-wave backgrounds
at $T_{\mathrm{kin}}\simeq100$--$110\,\mathrm{MeV}$. For a given background,
the measured polarization determines the geometric displacement
$\Delta\epsilon_{2,\mathrm{fo}}=\Pz/a$ from the cancellation boundary
$\efo=(b/a)v_2$. Longitudinal polarization therefore supplies geometric
information beyond the elliptic flow alone and can be combined with spectra
and femtoscopy to constrain the decoupling source.

\begin{acknowledgments}
The work of Y. L. Xie is supported by the National Natural Science Foundation
of China under Grant No.~12005196.
\end{acknowledgments}

\appendix

\section{Fitting and validation protocol}\label{app:fit}

The linear relation~(\ref{eq:linear}) is obtained by ordinary least squares
without intercept on the 48 calibration states, with $\efo$ and $v_2$ as the
two regressors: $a=0.010005$, $b=0.013631$, $\Cdet=0.866$, and 41 correct
signs. For Eq.~(\ref{eq:total}), the exponent $1/2$ is fixed, so that the
relation is linear in $C=AB$ and $D=-A$ with regressors
$v_2S_\epsilon(\efo/v_2)^{1/2}$ and $v_2S_\epsilon$; this gives $A=0.0404$ and
$B=0.850$. No state is removed because it lies close to the cancellation
boundary. The coefficient of determination is
$\Cdet=1-\sum_i(P_i-P_i^{\mathrm{pred}})^2/\sum_i(P_i-\bar P)^2$. Grouped
cross-validation refits both coefficients after withholding each complete
$e_{\mathrm{peak}}$, $\sigma_\perp$, or $b_0$ group and predicts the withheld
states; the quoted $\Cdet$ is computed from the pooled out-of-sample
predictions of all folds, not as an average over folds, so individual folds can
have negative $\Cdet$ while the pooled value is positive. The 12 additional states are predicted with the
coefficients fixed; they were calculated after the coefficients of
Eq.~(\ref{eq:total}) were fixed, whereas Eq.~(\ref{eq:linear}) was introduced
later and fitted on the 48 states only. The analyses of Sec.~\ref{sec:origin} require the
decoupling surfaces and use the 35 states with archived surfaces
($e_{\mathrm{peak}}=20$, $30$, $40\,\mathrm{GeV}/\mathrm{fm}^3$; the state
$b_0=0.77$, $e_{\mathrm{peak}}=30$, $\sigma_\perp=1.2\,\mathrm{fm}$ is also
missing). On these states the same procedure gives $a=0.0105$, $b=0.0145$, and
$\Cdet=0.75$. Bootstrap intervals, all cross-validation folds, the additional
and $b_0=0.685$ states, the decoupling-temperature and pressure-softening
checks, an alternative kinematic-shear-factored relation, and the complete
state tables are given in the Supplemental Material~\cite{supp}.

\section{Identities of the ILE integrand}\label{app:identities}

\paragraph{Time-derivative channel.} With the definitions of
Eq.~(\ref{eq:ile}), the $z$ component contains gradients with one time index in
the combination $\epsilon^{z0i\beta}p_\beta(\Xi_{0i}-\omega_{0i})$, with
$\Xi_{0i}-\omega_{0i}=\partial_iu_0$. The term $\partial_tu_0$ does not enter.
The channel $\partial_tu_i$ contributes $+\Pz^{(t)}$ to the shear part and
$-\Pz^{(t)}$ to the vorticity part; numerically the two cancel to better than
$10^{-9}$ (relative) for every state.

\paragraph{Two-term split.} Equation~(\ref{eq:split}) holds for on-shell $p$
and arbitrary $\partial_\alpha u_\beta$, with $\chi=p\cdot u$, spatial
derivatives at fixed laboratory time, and $\bm\Omega=\nabla\times\bm u$. To derive
it, let the indices $i,j$ run over the three spatial directions (unlike the
transverse indices of Sec.~\ref{sec:saddle}) and define $G_{ij}=\partial_iu^j$,
$\sigma_{ij}=\tfrac12(G_{ij}+G_{ji})$, and $\gamma=u^0$. With Eq.~(\ref{eq:Kdef})
and $A^0=-\bm p\cdot\bm\Omega/2$, and after the cancellation of the time-derivative
channel, the tensor contraction of Eq.~(\ref{eq:ile}) gives
\begin{align}
K={}&(\bm p\times\nabla\gamma)_z+\frac E2\,\Omega_z-\frac{[\bm p\times(\sigma\bm p)]_z}{E}\nonumber\\
&-\frac{p_z\,\bm p\cdot\bm\Omega}{2(E+m)},
\label{eq:Kraw}
\end{align}
where the last term comes from the transformation to the rest frame and must be
kept for a finite rapidity acceptance. With $\chi=E\gamma-\bm p\cdot\bm u$,
\begin{equation}
\nabla\chi=E\nabla\gamma-G\bm p,\qquad \sigma\bm p=G\bm p-\tfrac12\,\bm p\times\bm\Omega.
\end{equation}
Inserting these into Eq.~(\ref{eq:Kraw}) and using
$\bm p\times(\bm p\times\bm\Omega)=\bm p(\bm p\cdot\bm\Omega)-\bm p^2\bm\Omega$ and
$E^2-\bm p^2=m^2$ gives Eq.~(\ref{eq:split}). The derivation uses neither vanishing
T-vorticity nor the equations of motion nor a saddle-point approximation. As a
check of signs and implementation, Eqs.~(\ref{eq:split}) and (\ref{eq:Kraw}) agree
with the tensor contraction of Eq.~(\ref{eq:ile}) on 4096 random gradient sets and
on-shell momenta (maximum scaled difference $1.3\times10^{-15}$).

\paragraph{Curvature identity.} For a hypersurface $\Sigma$ in $(t,x,y,z)$ with
Euclidean unit normal $\hat n$, tangential gradient
$\nabla^\Sigma=\partial-\hat n(\hat n\cdot\partial)$, shape operator
$S_{\mu\nu}=\nabla^\Sigma_\mu\hat n_\nu$, $\kappa_\Sigma=\mathrm{tr}\,S$, boundary
$\partial\Sigma$ with two-dimensional area element $dS$ and outward co-normal
$\nu$, any smooth $\varphi$, and any constant vector $e$,
\begin{equation}
\int_\Sigma dA\,e\cdot\nabla^\Sigma\varphi
=\int_\Sigma dA\,\varphi\,\kappa_\Sigma\,(e\cdot\hat n)+\oint_{\partial\Sigma}dS\,\varphi\,(e\cdot\nu).
\end{equation}
With $f=e^{-\chi/T}$, constant $T$, and
$\nabla^\Sigma f=-f\nabla^\Sigma\chi/T$, the choice $\varphi=(p\cdot\hat n)f$ and
$e=\ell$ gives Eq.~(\ref{eq:curvature}); the term $\ell^\mu S_{\mu\nu}p^\nu$
arises from $\nabla^\Sigma(p\cdot\hat n)$. On a closed ellipsoid in four
dimensions with semi-axes $(3,4,2.5,3)\,\mathrm{fm}$, a prescribed smooth flow,
and $T=160\,\mathrm{MeV}$, sampled with $3\times10^5$--$9\times10^5$ random
points and $S_{\mu\nu}$ from a local linear fit of $\hat n$ over 32 neighbors,
the two sides agree to 0.5--7\% (about 4\% on average) over four random seeds.

\begin{table}[!htbp]
\caption{Approximate curvature form of Eq.~(\ref{eq:curvature}) on the 35 PICR
states for different numbers $k$ of neighboring facets in the fit of the shape
operator: number of states with a positive value, mean (in units of
$10^{-3}$), ratio of the mean to the mean tangential emission gradient
($0.87\times10^{-3}$), and correlation coefficients with the tangential
emission gradient, $\efo$, and $v_2$. The last two rows use a subset of 12
states (every third state), with a mean tangential emission gradient of
$0.97\times10^{-3}$.}
\label{tab:knn}
\centering\footnotesize
\setlength{\tabcolsep}{4pt}
\begin{tabular}{rcccccc}
\toprule
$k$ & positive & mean & ratio & $r(\text{tangential})$ & $r(\efo)$ & $r(v_2)$\\
\midrule
32 & 33/35 & 1.78 & 2.0 & 0.53 & 0.51 & 0.23\\
64 & 35/35 & 1.72 & 2.0 & 0.70 & 0.63 & 0.28\\
128 & 35/35 & 1.29 & 1.5 & 0.80 & 0.79 & 0.17\\
256 & 35/35 & 1.04 & 1.2 & 0.84 & 0.82 & 0.16\\
\midrule
256 & 12/12 & 1.11 & 1.14 & 0.89 & 0.87 & 0.17\\
512 & 12/12 & 1.02 & 1.05 & 0.91 & 0.90 & 0.19\\
\bottomrule
\end{tabular}
\end{table}

\section{Channel decomposition and derivative estimates}\label{app:channels}

\paragraph{Vanishing T-vorticity and the cooling rate.} In an ideal uncharged
fluid, an initially vanishing T-vorticity $d(Tu)$ is preserved by the
flow~\cite{becattini2018,becattini2015,denghuang2016}, and the velocity gradients
then satisfy
\begin{equation}
\partial_\mu u_\nu-\partial_\nu u_\mu=-(u_\nu\partial_\mu\ln T-u_\mu\partial_\nu\ln T).
\label{eq:irrot}
\end{equation}
For a boost-invariant flow decoupling at constant proper time with $T=T(\tau)$,
Eq.~(\ref{eq:irrot}) gives $P_z\propto(dT/d\tau)\,v_2/(mT)$~\cite{becattini2018}.
On the PICR decoupling surfaces, the computed velocity and temperature gradients
deviate from Eq.~(\ref{eq:irrot}) by an amount comparable to the kinematic
vorticity itself: inserted into Eq.~(\ref{eq:ile}), Eq.~(\ref{eq:irrot}) gives a
kinematic-vorticity contribution about twice the computed $\Pz^\omega$, with a
correlation of $0.75$ over the 35 states. A fit $|\Pz^\omega|=c\,\lambda v_2$ with
the cooling rate $\lambda$ of Sec.~\ref{sec:acc} gives $\Cdet=0.53$, compared with
$0.63$ for $c\,S_\epsilon v_2$ on the same states.

\paragraph{Gradient channels.} 
We integrated the vorticity and shear contributions separately for each of the
16 entries $\partial_\alpha u_\beta$ and for facets with time-like and
space-like normals, with the signed Cooper-Frye weight and acceptance on a
$(p_T,\phi,y)=11\times12\times3$ grid. The sums reproduce the tabulated
components to $0.07\times10^{-3}$. The tangential/normal split uses the
Euclidean normal. The tangential part was evaluated (i) with the stored lattice
gradients and (ii) with tangential velocity derivatives refitted from the facet
values of $u$ over the 32 nearest facets. Over the 35 states the two estimates
have means $0.87$ and $0.82\times10^{-3}$, an rms difference of
$0.34\times10^{-3}$, and a median absolute difference of 3\% (maximum 15\%) of
$|\Pz^\omega|+|\Pz^\Xi|$; for the state $b_0=0.77$,
$e_{\mathrm{peak}}=20\,\mathrm{GeV}/\mathrm{fm}^3$, $\sigma_\perp=0.8\,\mathrm{fm}$
they are $2.02$ and $2.64\times10^{-3}$. Both correlate with $\efo$ ($r=0.82$ and
$0.85$). Discretization, interpolation to the facets, and derivative reconstruction
can all contribute to this difference. Although their individual effects
have not been separated by a grid-convergence study, both estimates retain
the positive mean and the correlation with freeze-out eccentricity.

For the approximate curvature evaluation of Sec.~\ref{sec:curv}, the shape
operator of each facet is the tangential projection of a regularized linear
fit of the Euclidean unit normals over the $k$ nearest facets in $(t,x,y,z)$,
using only neighbors whose normals differ by less than $60^\circ$; the
boundary term of Eq.~(\ref{eq:curvature}) is neglected. Table~\ref{tab:knn}
shows how the result depends on $k$. Small neighborhoods are dominated by the
noise of the facet normals; over the tested values of $k$, the mean curvature estimate approaches
the mean tangential contribution computed from the stored velocity gradients,
and their correlation increases. Because $k=512$ is costly, it was evaluated only on every
third state (12 states); there the mean of the curvature form lies 5\% above
the tangential mean, or 10\% with the boundary term added.

The boundary term of Eq.~(\ref{eq:curvature}) is estimated from the facets
within one Cornelius time step ($\Delta t\simeq0.17\,\mathrm{fm}$) of the
first and of the last time slice. For these facets the outward co-normal is
$\nu=\mp\,\mathcal P e_t/|\mathcal P e_t|$, where $\mathcal P=1-\hat n\hat n$ projects onto the
surface and $e_t$ is the unit time vector, and the edge area is
$dS=dA\,|\mathcal P e_t|/\Delta t$. On a four-dimensional ellipsoid opened at a fixed
time, with a prescribed velocity field, this estimate reproduces the edge area
within 1\%, and the curvature form plus the boundary term reproduces the
tangential emission gradient within 1.3\%. On the 35 PICR states the boundary
term is $0.037\times10^{-3}$ on average ($0.030\times10^{-3}$ with a slab of two
time steps), at most $0.13\times10^{-3}$, and it is dominated by the initial
edge.

\section{Blast-wave model}\label{app:bw}

\paragraph{Fields and derivatives.} With the fields of Sec.~\ref{sec:bw} and
$u=(\gamma_\perp\cosh\eta_s,U^x,U^y,\gamma_\perp\sinh\eta_s)$,
$\gamma_\perp=(1+\bm U^2)^{1/2}$, the transverse derivatives at fixed $\tau$ are
$\partial_iU^j=(T_f/T)\partial_i\partial_j\Phi-U^j\partial_i\ln T$ and
$\partial_i\gamma_\perp=U^j\partial_iU^j/\gamma_\perp$. For the general profile
$\bm U=Q^{(n-1)/2}(H_xx,H_yy)$ on the surface, $Q=(H_xx^2+H_yy^2)/(Hr_0^2)$, the
potential is $\Phi=Hr_0^2Q^{(n+1)/2}/(n+1)$; $n=1$ is the reference linear
profile. At fixed laboratory time, $\partial_z=-\sinh\eta_s\,\partial_\tau$ on
functions of $(\tau,x,y)$, so
$\Omega_x=\sinh\eta_s(\partial_y\gamma_\perp+\partial_\tau U^y)$ and
$\Omega_y=-\sinh\eta_s(\partial_x\gamma_\perp+\partial_\tau U^x)$, with
$\partial_\tau U^j=-U^j\partial_\tau\ln T$. The algebraic curl check of Table~\ref{tab:bw}
replaces only these two components by the values implied by
$\partial_\tau(TU^j)=-\partial_j(T\gamma_\perp)$, i.e.,
$\bm\Omega_\perp=(\bm u\times\nabla\ln T)_\perp$, and keeps all other
quantities unchanged.

\paragraph{Closedness condition.} On the isotherm, $d(Tu)=0$ implies that the
pull-back of $u_\mu dx^\mu$ to $\Sigma$, $\gamma_\perp d\tau_s-\bm U\cdot d\bm x$,
is closed. With $\bm U=(H_xx,H_yy)$ on the surface this requires
$\partial_x\gamma_\perp\partial_y\tau_s=\partial_y\gamma_\perp\partial_x\tau_s$,
i.e., $H_x^2/r_y^2=H_y^2/r_x^2$, which at first order is
$\delta=\epsilon_R/2$, so that the two anisotropies could not be varied
independently. The time-space components of $d(Tu)$ enter the integrand only
through the transverse spatial-curl components $\Omega_{x,y}\propto\sinh\eta_s$;
replacing these components by the values implied by $d(Tu)=0$ (curl check of
Table~\ref{tab:bw}) changes $a$ by $+5\%$ and $b$ by $-10\%$ at the reference
point. The longitudinal component $\Omega_z$ in Eq.~(\ref{eq:bwcurl}) is the value
implied by the transverse condition $\nabla_\perp\times(T\bm U)=0$, whereas the PICR gradients do not quantitatively satisfy the estimate based on
vanishing T-vorticity (Appendix~\ref{app:channels}). To assess how the curl
normalization affects the coefficients, we compare the curl contributions: the
curl term gives 28\% of $b$ at the reference point,
comparable to 36\% on the PICR surfaces (Table~\ref{tab:curv}), and reducing
the blast-wave curl term by a factor of four lowers $b/a$ only from $1.43$ to
$1.28$.

\paragraph{Observables and numerics.} The surface element is
$d\Sigma_\mu=(\cosh\eta_s,-\partial_x\tau_s,-\partial_y\tau_s,-\sinh\eta_s)\,\tau_s\,d\eta_s\,d^2x$.
The polarization uses the integrand~(\ref{eq:split}) with the weight
$(p\cdot d\Sigma)\,e^{-(p\cdot u-m)/T_f}$. The acceptance is
$0.5<p_T<6\,\mathrm{GeV}/c$ (24 Gauss-Legendre nodes), $\phi$ (24 nodes), and
$|y|<0.5$ (3 nodes), with $\eta_s\in[-3,3]$ (24 nodes). The transverse grid uses
$\rho=\sin\vartheta$ with $48\times48$ nodes, which removes the rim singularity
of $\partial\tau_s$. $\efo$ uses the full $u^\mu d\Sigma_\mu$ weight. Refining the
grids changes $a$ and $b$ by at most 3\%.

\paragraph{Coefficients.} For each background, the responses of
$(\efo,v_2,\Pz)$ to $\epsilon_R$ and $\delta$ are obtained by central
differences with step $0.02$. The coefficients follow from $(a,-b)=J\,M^{-1}$,
where $M$ is the $2\times2$ response matrix of $(\efo,v_2)$ and $J$ that of
$\Pz$ (condition number about 2). Masses: $m_p=0.9383$, $m_\Lambda=1.1157$,
$m_\Xi=1.3217$, $m_\Omega=1.6725\,\mathrm{GeV}$, all with the spin-$\tfrac12$
integrand.

\paragraph{State-by-state mapping.} To test whether the model describes
individual PICR states, we read its parameters off each PICR surface without
using $\Pz$. $r_x$ and $r_y$ follow from the $u\cdot d\Sigma$-weighted second moments of
the midplane ($|z|<0.5\,\mathrm{fm}$) surface, $H_x$ and $H_y$ from a
weighted least-squares fit $U^j=H_jx^j$ of the flow on that surface, $\tau_0$
from the earliest decoupling time, $\Delta\tau$ from the latest decoupling time
at the center, and the central cooling rate from the $\Lambda$-yield-weighted
$\partial_t\ln T$. The archived times are elapsed times since the initial slice
$t_0=4\,\mathrm{fm}/c$. Adding this offset gives the physical decoupling
times and $\tau_0\simeq4.2\,\mathrm{fm}/c$. Because a PICR surface can
have several decoupling times at the same transverse position, we match
these moments and time scales instead of fitting a single graph $t(x,y)$.

With these parameters the model follows the state-by-state variation of $v_2$ ($r=0.99$, mean
22\% high) and the ordering of $\Pz$ ($r=0.83$, 26 of 35 signs), but its $\efo$
is only $0.35$ of the PICR value and the spread of its $\Pz$ is 1.6 times
larger.

If instead $\epsilon_R$ is chosen such that the model reproduces the
PICR $\efo$, $v_2$ remains correlated ($r=0.98$, mean 7\% low) and
$r(\Pz)=0.84$, but $\Pz$ is shifted by $+1.5\times10^{-3}$ and only 19 signs
agree. Two-variable fits to the mapped model give $b/a=0.62$ and $0.94$ in the
two variants. The mapped
parameters, $H=0.09$--$0.21\,\mathrm{fm}^{-1}$, $\Delta\tau=3.7$--$10.5\,\mathrm{fm}$,
and $r_0=4.1$--$7.3\,\mathrm{fm}$, lie in the direction in which Table~\ref{tab:bw}
shows $b/a$ to decrease. The ordering alone is not a sharp test, because the
PICR $\Pz$ correlates with $\efo-\xi v_2$ at $r>0.83$ for any
$0.6<\xi<1.6$.

To test the role of the start time, we recalculated the
six states with $e_{\mathrm{peak}}=20\,\mathrm{GeV}/\mathrm{fm}^3$,
$b_0=0.60$ and $0.77$, and $\sigma_\perp=0.8$, $1.1$, and $1.5\,\mathrm{fm}$,
starting at $t_0=2\,\mathrm{fm}/c$ with the same profile in $\eta_s$, the
Bjorken flow $v_z=z/t_0$, and $e_{\mathrm{peak}}$ multiplied by $2^{4/3}$. The
initial total energy is then 1.23 times the original one, close to the factor
$2^{1/3}=1.26$ expected from the longitudinal work in Bjorken expansion; with
$t_0=2\,\mathrm{fm}/c$ the fluid covers about 11 longitudinal cells instead of
22. Relative to the original states, the mapped flow strength rises by
19--25\%, $v_2$ by 9--22\%, and $r_0$ by 2--13\%, whereas $\efo$ falls by
17--20\%; $\Delta\tau$ decreases by up to 9\% in five states and by half in
one. Two-variable fits over the six states give $b/a=1.34$ (original) and
$1.05$ (earlier start) for PICR, $0.63$ and $0.88$ for the first mapping
variant, and $-0.06$ and $0.97$ for the second.

\end{document}


\title{Supplemental Material for ``Longitudinal \texorpdfstring{$\Lambda$}{Lambda}
polarization as a competition between freeze-out geometry and elliptic flow in
Pb-Pb collisions at
\texorpdfstring{$\sqrt{s_{\mathrm{NN}}}=5.36\,\mathrm{TeV}$}{sqrt(sNN) = 5.36 TeV}''}

\author{Yilong Xie}
\email{xieyl@cug.edu.cn}
\affiliation{School of Mathematics and Physics, China University of Geosciences (Wuhan), Lumo Road 388, 430074 Wuhan, China}
\author{Simin Wu}
\affiliation{School of Mathematics and Physics, China University of Geosciences (Wuhan), Lumo Road 388, 430074 Wuhan, China}
\author{Laszlo P. Csernai}
\affiliation{Institute of Physics and Technology, University of Bergen, Allegaten 55, 5007 Bergen, Norway}

\maketitle

This Supplemental Material provides the fit definitions, coefficient
uncertainties, and validation results for the total ILE harmonic $\Pz$
discussed in Sec.~III and Appendix~A of the main text. The validation includes
grouped cross-validation, 12 states outside the fit, and changes of the
decoupling temperature and pressure. The data applications compare the
geometries inferred with different empirical relations and freeze-out
backgrounds, followed by the pion femtoscopic eccentricity calculated on the
PICR surfaces. The final sections discuss the $J_y=0$ initial states and
provide the complete calibration tables.

\section{Fit protocol and sampling stability}\label{sec:sm-fit}

The main text uses two relations for the total ILE harmonic
$\Pz\equiv\Pz^{\mathrm{ILE}}=\Pz^\omega+\Pz^\Xi$, the sum of the negative
kinematic-vorticity contribution $\Pz^\omega$ and the positive kinematic-shear
contribution $\Pz^\Xi$. The linear relation in the freeze-out eccentricity
$\efo$ and the elliptic flow $v_2$,
\begin{equation}
\Pz\simeq a\,\efo-b\,v_2,
\label{eq:linear-sm}
\end{equation}
is used to analyze the polarization contributions and the coefficients in
Secs.~IV and V of the main text, and to interpret the geometric information
in the ALICE inversion in Sec.~VI. The compact relation
\begin{equation}
\Pz\simeq
A v_2S_\epsilon
\left[
B\left(\frac{\efo}{v_2}\right)^{1/2}-1
\right],
\qquad
S_\epsilon=\frac{\efo}{\ein},
\label{eq:total-sm}
\end{equation}
was guided by the exact identity $\Pz=|\Pz^\omega|(R-1)$ with the competition
ratio $R=|\Pz^\Xi/\Pz^\omega|$, and it also contains the initial eccentricity
$\ein$ through the eccentricity survival fraction $S_\epsilon$. Both relations
use the polarization harmonic and $v_2$ in the same acceptance, $|y|<0.5$ and
$0.5<p_T<6\,\mathrm{GeV}/c$, with the $\Lambda$ yield as weight. The
freeze-out eccentricity $\efo$ is defined with the signed fluid-flux weight
$u^\mu d\Sigma_\mu$ on the decoupling surface.

Equation~(\ref{eq:linear-sm}) is fitted by ordinary least squares without an
intercept, with $\efo$ and $v_2$ as the two regressors. The 48 calibration
states give $a=0.010005$ and $b=0.013631$, with coefficient of determination
$\Cdet=0.866$, and Eq.~(\ref{eq:linear-sm}) vanishes at $\efo/v_2=b/a=1.362$.
For the fixed exponent $1/2$, Eq.~(\ref{eq:total-sm}) is linear in the two
coefficients $C=AB$ and $D=-A$,
\begin{equation}
\Pz\simeq
C v_2S_\epsilon
\left(\frac{\efo}{v_2}\right)^{1/2}
+D v_2S_\epsilon,
\label{eq:linear-fit}
\end{equation}
and the same least-squares procedure gives $A=0.0404$ and $B=0.8502$, with
$\Cdet=0.865$ and a cancellation boundary $\efo/v_2=1/B^2=1.383$. In both fits
no state is removed because it lies close to the cancellation boundary.

To measure how strongly the coefficients depend on the choice of calibration
states, we draw 10,000 bootstrap samples of 48 states with replacement and
repeat each two-coefficient fit. Table~\ref{tab:bootstrap} gives the resulting
percentile intervals. The amplitudes $a$, $b$, and $A$ vary by 13--19\%, whereas the
cancellation boundary is more stable: $b/a$ stays within $1.30$--$1.43$ and
$1/B^2$ within $1.30$--$1.47$. The ratio of the two coefficients, which decides
the sign of $\Pz$, is therefore the best-determined property of both relations.

\input{supplement_tables/fit_stability.tex}

\section{Grouped cross-validation and additional states}\label{sec:sm-cv}

The 48 calibration states combine four values of the initial energy-density
normalization $e_{\mathrm{peak}}$, six values of the transverse smoothing width
$\sigma_\perp$, and two values of the reduced impact parameter $b_0$. Grouped
cross-validation tests whether the relations depend on any single value of
these parameters. For each fold, one complete parameter group is withheld, both
coefficients are refitted to the remaining states, and the withheld states are
predicted. Table~\ref{tab:cv-summary} reports the pooled metrics, and
Table~\ref{tab:cv-folds} gives every fold. The linear relation,
Eq.~(\ref{eq:linear-sm}), and the compact relation, Eq.~(\ref{eq:total-sm}),
reach similar pooled values of $\Cdet$: $0.86$ and $0.81$ when one
$e_{\mathrm{peak}}$ group is withheld, $0.83$ and $0.84$ when one $\sigma_\perp$
group is withheld, and $0.74$ and $0.72$ when one $b_0$ group is withheld.

To identify why withholding a $b_0$ group gives the lowest pooled accuracy,
we compare the refitted cancellation points. When an
$e_{\mathrm{peak}}$ or a $\sigma_\perp$ group is withheld, the refitted $b/a$
stays between $1.34$ and $1.41$ (Table~\ref{tab:cv-folds}). When one $b_0$
group is withheld, only one transverse geometry remains, and the refitted
cancellation boundary moves to $b/a=1.22$ (fit on $b_0=0.77$) or $1.51$ (fit on
$b_0=0.60$); the compact relation behaves in the same way. Each $b_0$ fold must
therefore extrapolate from one transverse geometry to the other, and the
predicted signs of the states near the cancellation boundary change. The
isolated $e_{\mathrm{peak}}=10\,\mathrm{GeV}/\mathrm{fm}^3$ fold has a small
$\Cdet$ ($0.13$ for the linear relation and $-0.30$ for the compact relation)
but retains all 12 signs; extrapolation to the lowest energy-density
normalization is thus weaker for the magnitude of $\Pz$ than for its sign.

\input{supplement_tables/cv_summary.tex}
\input{supplement_tables/cv_folds.tex}

\begin{figure}[H]
\centering
\includegraphics[width=\textwidth]{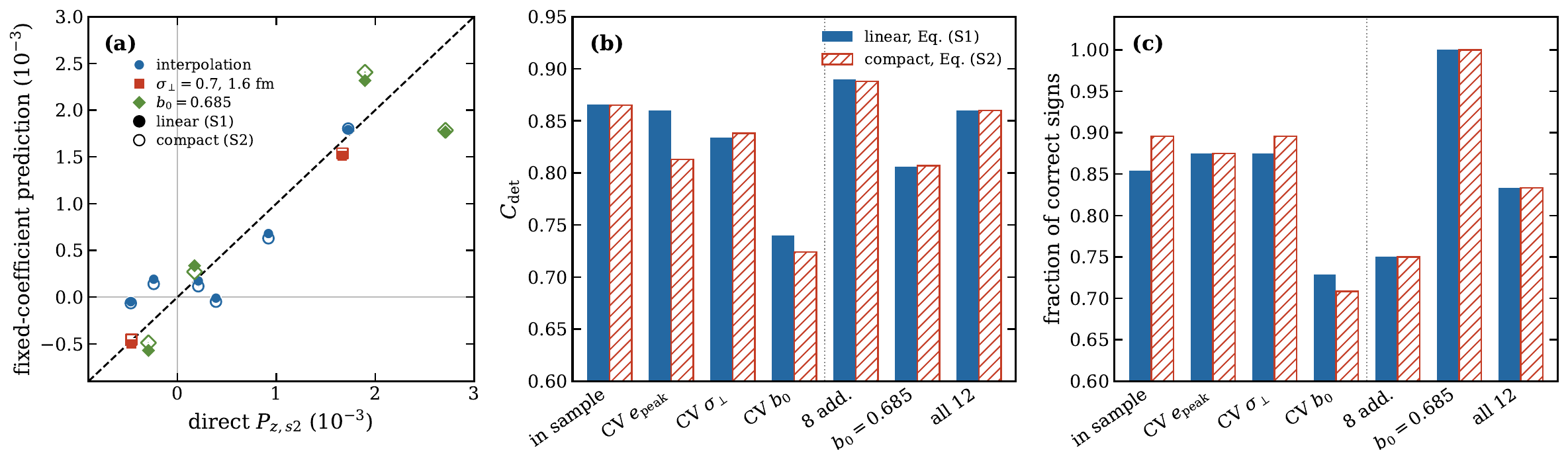}
\caption{(a) Predictions of the linear relation, Eq.~(\ref{eq:linear-sm})
(filled symbols), and the compact relation, Eq.~(\ref{eq:total-sm}) (open
symbols), with coefficients fixed on the 48 calibration states, versus the
direct calculation for the 12 additional states: six states interpolating in
$e_{\mathrm{peak}}$ and $\sigma_\perp$ (blue circles), two states with
$\sigma_\perp=0.7$ and $1.6\,\mathrm{fm}$ (red squares), and four states at
the previously unsampled $b_0=0.685$ (green diamonds). The dashed diagonal
denotes equality. (b) Coefficient of determination $\Cdet$ and (c) fraction
of correct signs for the in-sample fit, the three grouped cross-validations
(pooled out-of-sample predictions), the eight additional states, the four
states at $b_0=0.685$, and all 12 additional states. Solid bars refer to
Eq.~(\ref{eq:linear-sm}) and hatched bars to Eq.~(\ref{eq:total-sm}); the
dotted line separates the tests on the calibration states from those on the
additional states (Table~\ref{tab:cv-summary}).}
\label{fig:prospective}
\end{figure}

The test with additional states was defined before the eight additional
hydrodynamic calculations, with $A$ and $B$ of Eq.~(\ref{eq:total-sm}) fixed.
Six of these states interpolate within both the $e_{\mathrm{peak}}$ range and
the $\sigma_\perp$ range of the calibration set, and two extend the smoothing
range from $0.8$--$1.5\,\mathrm{fm}$ to $0.7$ and $1.6\,\mathrm{fm}$. Because
the $b_0$ folds require the largest extrapolation, we subsequently calculated
four states at the previously unsampled $b_0=0.685$, with
$e_{\mathrm{peak}}=10$ and $30\,\mathrm{GeV}/\mathrm{fm}^3$ and
$\sigma_\perp=0.8$ and $1.5\,\mathrm{fm}$. The linear relation,
Eq.~(\ref{eq:linear-sm}), was introduced after these calculations but fitted on
the 48 calibration states only, so none of the 12 additional states enters
either fit. Tables~\ref{tab:prospective} and \ref{tab:b0-midpoint} list the
direct calculations and both predictions, and Fig.~\ref{fig:prospective}
compares them.

For the eight additional states, Eq.~(\ref{eq:linear-sm}) gives $\Cdet=0.89$, a
root-mean-square error (RMSE) of $0.28\times10^{-3}$, and six correct signs;
Eq.~(\ref{eq:total-sm}) gives the same $\Cdet$ and RMSE to two digits. The two sign
misses are the same for both relations, V03 and V04, whose direct total
harmonics have magnitudes of only $0.24\times10^{-3}$ and $0.39\times10^{-3}$.
In both states the negative kinematic-vorticity contribution $\Pz^\omega$ and
the positive kinematic-shear contribution $\Pz^\Xi$ are much larger than their
sum, so a small error in either contribution reverses the sign of the
near-cancelled total. For the four states at $b_0=0.685$, both relations give
$\Cdet=0.81$, an RMSE of $0.54\times10^{-3}$, and all four signs correctly.
The direct total harmonics of these four states differ from the arithmetic
midpoint of the matching $b_0=0.60$ and $0.77$ states by only
$0.03$--$0.21\times10^{-3}$, so the total harmonic interpolates smoothly
between the two endpoint geometries. Over all 12 additional states, both
relations describe 86\% of the variation of $\Pz$ and predict 10 signs
correctly.

\input{supplement_tables/prospective_states.tex}

\input{supplement_tables/b0_midpoint_states.tex}

\section{Fixed-coefficient checks under decoupling and pressure softening}\label{sec:sm-checks}

Both relations are calibrated at $T_{\mathrm{dec}}=160\,\mathrm{MeV}$ with the
three-flavor bag-model equation of state of the main text. To see how the
fixed relations respond to a change of the decoupling condition, we change the
decoupling temperature at the common initial state
$e_{\mathrm{peak}}=10\,\mathrm{GeV}/\mathrm{fm}^3$,
$\sigma_\perp=1.5\,\mathrm{fm}$, and $b_0=0.77$, without refitting the
coefficients. In a second check at the same initial state and
$T_{\mathrm{dec}}=160\,\mathrm{MeV}$, the bag-model pressure
$p_{\mathrm{bag}}=(e-4B_{\mathrm{bag}})/3$ is replaced by
\begin{equation}
p_{\mathrm{soft}}(T)=p_{\mathrm{bag}}
\left\{1-0.40\exp\left[-\frac{1}{2}
\left(\frac{T-165\,\mathrm{MeV}}{30\,\mathrm{MeV}}\right)^2\right]\right\},
\label{eq:pressure-softening}
\end{equation}
while the bag-model relation between energy density and temperature is kept.
The temperature-dependent factor suppresses the pressure itself by up to
40\%; the pressure gradient also depends on the derivative of this factor.
Although this prescription is not a thermodynamically consistent crossover
equation of state, it tests how a change of the pressure law near decoupling
shifts the balance between $\Pz^\omega$ and $\Pz^\Xi$.

\begin{figure}[H]
\centering
\includegraphics[width=0.88\textwidth]{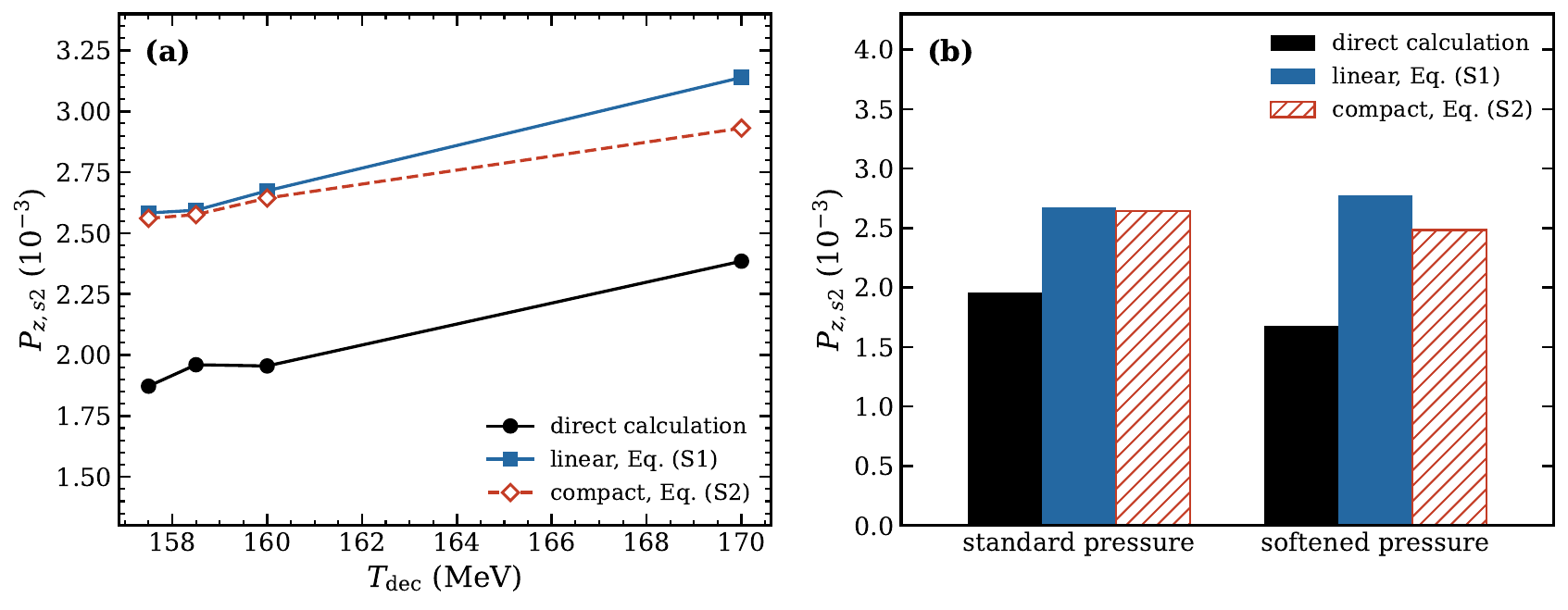}
\caption{Direct total ILE harmonic $\Pz$ (black) and the predictions of the
linear relation, Eq.~(\ref{eq:linear-sm}) (blue), and the compact relation,
Eq.~(\ref{eq:total-sm}) (red), with coefficients fixed on the 48 calibration
states, at the initial state
$e_{\mathrm{peak}}=10\,\mathrm{GeV}/\mathrm{fm}^3$,
$\sigma_\perp=1.5\,\mathrm{fm}$, $b_0=0.77$. (a) Decoupling-temperature scan
with the bag-model equation of state. (b) Standard and softened pressure at
$T_{\mathrm{dec}}=160\,\mathrm{MeV}$; the softened pressure is defined by
Eq.~(\ref{eq:pressure-softening}). The numerical values are given in
Table~\ref{tab:model-checks}.}
\label{fig:model-checks}
\end{figure}

Table~\ref{tab:model-checks} and Fig.~\ref{fig:model-checks} show the results.
The direct total harmonic remains positive at all four decoupling temperatures
and under the pressure softening, and both relations predict the correct sign
in every case. In the temperature scan, the linear relation overpredicts the
direct total by $0.63$--$0.75\times10^{-3}$ and the compact relation by
$0.55$--$0.69\times10^{-3}$, that is, by $11$--$16\%$ of the mean magnitude of
the two cancelling contributions, $(|\Pz^\omega|+|\Pz^\Xi|)/2$. This common
positive offset is already present at the $160\,\mathrm{MeV}$ reference and is
not generated by the change of the decoupling temperature. Relative to
$T_{\mathrm{dec}}=160\,\mathrm{MeV}$, the direct total changes by $-4\%$ to
$+22\%$, the linear relation by $-3\%$ to $+17\%$, and the compact relation by
$-3\%$ to $+11\%$, so both relations follow the increase of $\Pz$ toward
higher decoupling temperature.

The pressure softening reduces the direct total harmonic by $14\%$. The
compact relation predicts a reduction by $6\%$, whereas the linear relation
predicts a slight increase by $4\%$: the softening lowers $v_2$ from $0.058$ to
$0.046$ but changes $\efo$ by only 2\%, so the term $-b\,v_2$ becomes less
negative. The deviations from the direct total are $32\%$ and $44\%$ of the mean
component magnitude for the compact and the linear relation, respectively.
Although the fixed coefficients do not accurately reproduce the magnitude
of the response to pressure softening, both relations retain the positive,
shear-dominated sign. This single-state test does not determine a new pair
of coefficients, but it shows why the original calibration cannot be
transferred quantitatively to a modified pressure. The surviving sign
supports the vorticity-shear competition as the physical interpretation of
the total harmonic in this check.

\input{supplement_tables/model_checks.tex}

\section{Alternative kinematic-shear-factored relation and geometric inversion}\label{sec:sm-shear}

The compact relation, Eq.~(\ref{eq:total-sm}), uses a prefactor
$Av_2S_\epsilon$ guided by the form of the magnitude of the kinematic-vorticity
contribution $|\Pz^\omega|$. To check whether the cancellation boundary and the
freeze-out geometry inferred from data depend on this choice, we start instead
from the exact identity $\Pz=\Pz^\Xi(1-1/R)$ with
$R=|\Pz^\Xi/\Pz^\omega|$. A two-coefficient form guided by the positive
kinematic-shear contribution $\Pz^\Xi$ is
\begin{equation}
\Pz\simeq
\widetilde A\,\efo
\left[1-\widetilde B
\left(\frac{v_2}{\efo}\right)^{1/2}\right].
\label{eq:shear-factored}
\end{equation}
For the fixed exponent $1/2$, we fit the 48 direct total harmonics by ordinary
least squares without an intercept, with $\efo$ and $\sqrt{\efo v_2}$ as the
two regressors. The resulting coefficients are $\widetilde A=0.0162$ and
$\widetilde B=1.167$. Like $a$ and $b$, they are fitted to the total harmonic,
so $\widetilde A\efo$ is not a separate fit of $\Pz^\Xi$.

Table~\ref{tab:shear-comparison} compares the three relations on the 48
calibration states, in grouped cross-validation, and on the 12 additional
states. The linear and the compact relation have nearly equal accuracy, and
the kinematic-shear-factored relation has a slightly lower $\Cdet$ in every
comparison; it still describes the 12 additional states with $\Cdet=0.83$ and
predicts the same 10 signs, missing only the near-cancelled states V03 and V04.
The three relations differ in form but share their cancellation boundary:
Eq.~(\ref{eq:shear-factored}) changes sign at
$\efo/v_2=\widetilde B^2=1.362$, equal to $b/a=1.362$ of
Eq.~(\ref{eq:linear-sm}) and close to $1/B^2=1.383$ of
Eq.~(\ref{eq:total-sm}). Although the fitted amplitudes depend on the empirical form, all three
forms tested locate the sign change at nearly the same $\efo/v_2$. This
agreement supports the cancellation boundary as a stable feature of the
scanned states.

\input{supplement_tables/shear_factored_comparison.tex}

The same three relations can be inverted for the ALICE polarization harmonics
and the spectrum-weighted $\Lambda$ elliptic flow $v_2$ used in Sec.~VI A of
the main text. For a positive measured harmonic $\Pz^{\mathrm{exp}}$, the
linear relation gives $\efo=(\Pz^{\mathrm{exp}}+b\,v_2)/a$, and the
shear-dominated solution of Eq.~(\ref{eq:shear-factored}) is
\begin{equation}
\efo=
\frac{1}{4}\left[
\widetilde B\sqrt{v_2}+
\sqrt{\widetilde B^2v_2+
4\Pz^{\mathrm{exp}}/\widetilde A}
\right]^2.
\label{eq:shear-inversion}
\end{equation}
Table~\ref{tab:shear-inversion} compares the inferred freeze-out
eccentricities. At $5.36\,\mathrm{TeV}$, the linear relation gives
$\efo=0.208$ in 30--40\% and $0.274$ in 40--50\% collisions,
Eq.~(\ref{eq:shear-inversion}) gives $0.215$ and $0.284$, and the compact
relation gives the intervals $0.216$--$0.233$ and $0.270$--$0.305$ when the
initial eccentricity $\ein$ is varied over the envelope of the main text. All
values lie inside the 48-state calibration range $0.169<\efo<0.467$. The
linear and kinematic-shear-factored relations, which contain no $\ein$,
give values within or slightly below the $\efo$ range obtained by varying
$\ein$ in the compact relation. For these measurement inputs and globally
calibrated coefficients, the inferred geometry therefore changes little
among the three empirical forms.

\input{supplement_tables/shear_factored_inversion.tex}

Measurements that change $v_2$ by event-shape selection offer a further test of
the competition. A STAR event-shape-engineering measurement in Au+Au collisions
at $\sqrt{s_{\mathrm{NN}}}=19.6\,\mathrm{GeV}$ reports a positive
$P_{z,2}$-$v_2$ slope and a negative extrapolated linear-fit intercept for a
TPC-based analysis combination~\cite{kondo2026sm}. An event-shape-selected
slope is not the coefficient of $v_2$ at fixed freeze-out geometry, because
$\efo$ changes together with $v_2$ along the selected sequence of events; with
fixed coefficients,
$d\Pz/dv_2=a\,d\efo/dv_2-b$ can be positive even though $-b$ is negative.
Although the STAR collision energy and event-shape selection differ from our
Pb-Pb scan at $\sqrt{s_{\mathrm{NN}}}=5.36\,\mathrm{TeV}$, measurements of this
kind, combined with calculations of $\efo$ for the selected events, could
separate the dependence of $\Pz$ on the freeze-out geometry from its dependence
on the elliptic flow.

\section{Background dependence of the ALICE inversion}\label{sec:sm-background}

The inversions above use coefficients fitted over the whole PICR scan. Because
$\efo=\Pz/a+(b/a)v_2$, the inferred freeze-out geometry depends on both the
polarization scale $a$ and the cancellation point $b/a$, and both depend on the
freeze-out background (Sec.~V of the main text). This section estimates how
strongly the choice of background changes the inferred $\efo$.

\emph{Local PICR calibration.} For each centrality class we calculated PICR
states at the reduced impact parameter matched by $c\simeq b_0^2$:
$b_0=0.60$ for 30--40\% and $b_0=0.685$ for 40--50\%. Larger transverse
smoothing widths, $\sigma_\perp=1.5$--$2.1\,\mathrm{fm}$, bring the $\Lambda$
elliptic flow $v_2$ down to the measured values. Each background uses two
normalizations, $e_{\mathrm{peak}}=20$ and $40\,\mathrm{GeV}/\mathrm{fm}^3$,
and is evaluated with the original start time $t_0=4\,\mathrm{fm}/c$ and with
$t_0=2\,\mathrm{fm}/c$, for which $e_{\mathrm{peak}}$ is multiplied by
$2^{4/3}$ to match the original energy-density scale at $4\,\mathrm{fm}/c$ under
the ideal Bjorken-cooling estimate. The evolution runs for 400 time steps
($16.7\,\mathrm{fm}/c$), so that the decoupling surfaces close also for the
longest-lived states. Table~\ref{tab:local-calib} lists the six-state fits of
Eq.~(\ref{eq:linear-sm}) and the resulting inversion. The uncertainty of the
inverted $\efo$ includes the coefficient uncertainty with the correlation of
$a$ and $b$, propagated with a common polarization uncertainty for each six-state fit,
equal to the larger of its residual standard error (four residual degrees of
freedom) and the numerical uncertainty $0.17\times10^{-3}$.

Table~\ref{tab:local-calib} shows that the local fits describe their six
states with $\Cdet=0.83$--$0.99$.
The start time changes both coefficients: $t_0=2\,\mathrm{fm}/c$ develops more
transverse flow and lowers the cancellation point from $b/a\simeq1.6$ to
$1.3$. At $5.36\,\mathrm{TeV}$, the inverted $\efo$ falls from $0.27$ to $0.19$
in 30--40\% and from $0.33$ to $0.25$ in 40--50\% collisions; about half of the
change comes from the term $\Pz/a$ and half from $(b/a)v_2$. These differences
exceed the 68\% intervals of each calibration. Although the local PICR samples underpredict the measured $\Lambda$ mean
transverse momentum by 10--30\% and the inversion can extend up to 0.08
beyond their eccentricity range, the local calibrations bracket the global
inversion of the main text. They therefore quantify a substantial
background dependence even when the calibration states span the measured
elliptic flow.

\input{supplement_tables/local_calibration.tex}

\emph{Blast-wave background from measured spectra.} As a background that does
not rely on PICR, we use the blast-wave model of Sec.~V and Appendix~D of the
main text with the radial flow of the ALICE blast-wave fit to identified-hadron
spectra~\cite{alicebwsm}. The model adopts the fitted velocity profile
$\beta_T(r)=\beta_s(r/r_0)^n$ with the fitted exponent $n$ and surface velocity
$\beta_s=\langle\beta_T\rangle(n+2)/2$, which reproduces the fitted mean
velocity $\langle\beta_T\rangle$ exactly. The coefficients follow from the
linear response to the spatial and flow anisotropies, so the anisotropy
parameters of a fit are not needed. The spectra do not fix the source radius
$r_0$ and the time depth $\Delta\tau$ of the decoupling surface, which we vary over
$r_0=8$--$12\,\mathrm{fm}$ and $\Delta\tau=2$--$6\,\mathrm{fm}$. The central
decoupling proper time is fixed at $9\,\mathrm{fm}/c$ by setting
$\tau_0=9\,\mathrm{fm}/c-\Delta\tau$ and $\tau_c=9\,\mathrm{fm}/c$. At $T_f=160\,\mathrm{MeV}$, this flow gives a $\Lambda$ mean transverse
momentum 18--26\% above the measurement. We therefore set the ILE emission temperature to the fitted kinetic
freeze-out temperature, $T_f=T_{\mathrm{kin}}$ (Table~\ref{tab:bw-alice}),
in the spectrum-constrained background. Although the model omits resonance decays, at
$T_f=T_{\mathrm{kin}}$ the calculated mean in the measured window,
$0.6<p_T<6\,\mathrm{GeV}/c$, lies only 5--10\% above the measured value.

Table~\ref{tab:bw-alice} shows the result. The cancellation point
$b/a=1.26$--$1.65$ covers the PICR values, and its median, about 1.4, is close
to the global PICR value 1.36. The scale $a$ varies by a factor of three with
$r_0$ and $\Delta\tau$, but the inverted $\efo$ varies less, because the term
$(b/a)v_2$ dominates. At $5.36\,\mathrm{TeV}$ the inverted $\efo$ is
$0.18$--$0.28$ (median $0.23$) in 30--40\% and $0.23$--$0.39$ (median $0.30$)
in 40--50\% collisions. The blast-wave model uses the same ILE integrand as the
PICR calculation, so it replaces the background but not the polarization
prescription. The quoted ranges describe variation over the twelve backgrounds at fixed
central measurement values. Although the source radius and time depth are not
fixed by the spectra and leave a spread of about $\pm25\%$, the model with
spectrum-constrained radial flow gives eccentricities overlapping those
from PICR, with medians close to the global inversion.

\input{supplement_tables/blastwave_alice.tex}

\section{Pion femtoscopic eccentricity on the PICR surfaces}\label{sec:sm-hbt}

Section VI of the main text compares the inferred $\efo$ with the femtoscopic
eccentricity measured by ALICE. To see how the two eccentricities are related
in the model, we evaluate the same femtoscopic estimator on the 35 PICR
surfaces. We use the emission function
$w(x,K)=\max(K\cdot d\Sigma,0)\,f_{\mathrm{BE}}(K\cdot u/T)$ for direct thermal
charged pions ($m_\pi=0.13957\,\mathrm{GeV}$, $\mu_\pi=0$, no resonance decays).
In the longitudinally comoving system (LCMS) at pair rapidity $Y=0$ and pair
azimuth $\Phi$, with $x_{\mathrm{out}}=x\cos\Phi+y\sin\Phi$ and
$x_{\mathrm{side}}=-x\sin\Phi+y\cos\Phi$, the Gaussian radii are the
space-time variances
$R_{\mathrm{side}}^2=\mathrm{Var}(x_{\mathrm{side}})$ and
$R_{\mathrm{out}}^2=\mathrm{Var}(x_{\mathrm{out}}-\beta_Tt)$, with
$\beta_T=k_T/\sqrt{m_\pi^2+k_T^2}$. Each $k_T$ is treated separately with 24
values of $\Phi$. Writing
$R_{\mathrm{side}}^2(\Phi)=R_{s,0}^2+2R_{s,2}^2\cos2\Phi+\cdots$, we define
$\epsilon_{\mathrm{HBT}}(k_T)=2R_{s,2}^2/R_{s,0}^2$. For $k_T\to0$ the weight
remains $f_{\mathrm{BE}}(m_\pi u^0/T)$ times the flux, so flow and the pion mass
still affect the homogeneity region. The side radii are $1.7$--$3.0\,\mathrm{fm}$,
and the ALICE ranges cover the $k_T$ bins $0.2$--$0.3$, $0.3$--$0.4$,
$0.4$--$0.5$, and $0.5$--$0.7\,\mathrm{GeV}/c$.

Table~\ref{tab:hbt} shows the result. At $k_T=0.25\,\mathrm{GeV}/c$ the
femtoscopic eccentricity is on average 0.91 of $\efo$, with a state-by-state
correlation of 0.70. The two eccentricities sample different emission
weights: positive-flux Bose-Einstein weights at a selected pair
momentum versus the fluid-flux weight of the whole surface. It retains
0.39--0.63 of $\ein$, compared with 0.2--0.4 in the ALICE data. When
$\epsilon_{\mathrm{HBT}}$ replaces $\efo$ in Eq.~(\ref{eq:linear-sm}), with both
coefficients refitted, $C_{\mathrm{det}}$ is 0.53 at $k_T=0.25\,\mathrm{GeV}/c$
and decreases with increasing $k_T$; it rises to 0.72 for $k_T\to0$, compared
with 0.75 for $\efo$. At the pair momenta of the measurement, the pion
femtoscopic eccentricity is therefore a less effective variable than $\efo$ in
the polarization relation.

\input{supplement_tables/pion_hbt.tex}

\section{Scope of the \texorpdfstring{$J_y=0$}{Jy=0} initial states}\label{sec:sm-jy}

The reflection-symmetric Bjorken longitudinal profile sets the total angular
momentum of the hydrodynamic initial state to $J_y=0$. This condition does not
set the local transverse velocity gradients to zero: anisotropic expansion
still generates the kinematic-vorticity contribution $\Pz^\omega$ and the
kinematic-shear contribution $\Pz^\Xi$ studied in the main text. The $J_y=0$
construction therefore isolates how the initial transverse energy-density
geometry controls the competition between these two local contributions,
without simultaneously varying a longitudinal tilt or a global rotation.

A tilted initial energy-density profile with nonzero $J_y$ adds longitudinal
flow and vorticity. Such longitudinal structure is important for global
polarization and directed flow~\cite{ryu2021sm}, and event-by-event
three-dimensional calculations show that longitudinal-polarization harmonics
also respond to the fluctuating flow field and to hydrodynamic
parameters~\cite{alzhrani2022sm}. Nonzero $J_y$ can therefore modify $\Pz^\omega$
and $\Pz^\Xi$ through the velocity field and the decoupling hypersurface, and
with them the coefficients $a$ and $b$ of Eq.~(\ref{eq:linear-sm}). Because the
exact identities of the ILE integrand used in Sec.~IV of the main text hold for
any velocity field, the split of $\Pz$ into the emission-gradient and curl
terms also applies to tilted initial states; a calculation with tilted
event-by-event initial conditions would determine how the coefficients $a$ and
$b$ change.

\section{Complete calibration table}\label{sec:sm-tables}

Tables~\ref{tab:training-states} and \ref{tab:training-harmonics} list the 48
calibration states, their direct polarization harmonics, and the fitted total
harmonics of Eqs.~(\ref{eq:linear-sm}) and (\ref{eq:total-sm}). They provide
the numerical inputs needed to reproduce both fits. The linear relation misses
the sign of seven states (13, 29, 30, 32, 33, 46, and 47) and the compact
relation of five (13, 30, 32, 46, and 47). All of these states have
$\efo/v_2=1.25$--$1.48$, close to the cancellation boundary
$\efo/v_2\simeq1.36$--$1.38$, and all have $|\Pz|<0.4\times10^{-3}$.

\input{supplement_tables/training_states.tex}

%% file: supplement_tables/fit_stability.tex
\begin{table}[H]
\caption{Sampling stability of the fitted coefficients and of the cancellation
boundary for the linear relation, Eq.~(\ref{eq:linear-sm}), and the compact
relation, Eq.~(\ref{eq:total-sm}). The interval is the 2.5--97.5 percentile
range from 10,000 bootstrap resamples of the 48 calibration states. The
coefficients $a$, $b$, and $A$ are given in units of $10^{-3}$.}
\label{tab:bootstrap}
\centering
\begin{tabular}{llccc}
\toprule
Relation & Quantity & Full fit & Bootstrap median & 95\% interval \\
\midrule
Linear, Eq.~(\ref{eq:linear-sm}) & $a$ & 10.01 & 9.99 & 8.69--11.68 \\
 & $b$ & 13.63 & 13.63 & 11.68--16.13 \\
 & $b/a$ & 1.362 & 1.365 & 1.296--1.433 \\
\addlinespace
Compact, Eq.~(\ref{eq:total-sm}) & $A$ & 40.4 & 40.4 & 34.7--48.1 \\
 & $B$ & 0.8502 & 0.8491 & 0.8251--0.8780 \\
 & $1/B^2$ & 1.383 & 1.387 & 1.297--1.469 \\
\bottomrule
\end{tabular}
\end{table}

%% file: supplement_tables/cv_summary.tex
\begin{table}[H]
\caption{Aggregate validation metrics of the linear relation,
Eq.~(\ref{eq:linear-sm}), and the compact relation, Eq.~(\ref{eq:total-sm}).
For grouped cross-validation, $C_{\mathrm{det}}$ and the root-mean-square
error (RMSE) are computed from the pooled out-of-sample predictions of all
folds. For the additional states, both coefficients of each relation are fixed
at their 48-state values. The RMSE is given in units of $10^{-3}$.}
\label{tab:cv-summary}
\centering
\begin{tabular}{lcccccc}
\toprule
& \multicolumn{3}{c}{Linear, Eq.~(\ref{eq:linear-sm})}
& \multicolumn{3}{c}{Compact, Eq.~(\ref{eq:total-sm})} \\
\cmidrule(lr){2-4}\cmidrule(lr){5-7}
Test & $C_{\mathrm{det}}$ & RMSE & Correct sign & $C_{\mathrm{det}}$ & RMSE & Correct sign \\
\midrule
48-state fit & 0.87 & 0.41 & 41/48 & 0.87 & 0.41 & 43/48 \\
Leave one $e_{\mathrm{peak}}$ group out & 0.86 & 0.42 & 42/48 & 0.81 & 0.48 & 42/48 \\
Leave one $\sigma_\perp$ group out & 0.83 & 0.45 & 42/48 & 0.84 & 0.45 & 43/48 \\
Leave one $b_0$ group out & 0.74 & 0.57 & 35/48 & 0.72 & 0.59 & 34/48 \\
Eight additional states & 0.89 & 0.28 & 6/8 & 0.89 & 0.28 & 6/8 \\
Four states at $b_0=0.685$ & 0.81 & 0.54 & 4/4 & 0.81 & 0.54 & 4/4 \\
All 12 additional states & 0.86 & 0.39 & 10/12 & 0.86 & 0.39 & 10/12 \\
\bottomrule
\end{tabular}
\end{table}

%% file: supplement_tables/cv_folds.tex
\begin{table}[H]
\caption{Fold-by-fold grouped cross-validation. Both coefficients of each
relation are refitted after removing the complete group shown in the first two
columns and are then applied only to that withheld group. The coefficients $a$,
$b$, and $A$ are given in units of $10^{-3}$.}
\label{tab:cv-folds}
\centering
\resizebox{\textwidth}{!}{%
\begin{tabular}{lccccccccccc}
\toprule
& & \multicolumn{5}{c}{Linear, Eq.~(\ref{eq:linear-sm})}
& \multicolumn{5}{c}{Compact, Eq.~(\ref{eq:total-sm})} \\
\cmidrule(lr){3-7}\cmidrule(lr){8-12}
Withheld axis & Value & $a$ & $b$ & $b/a$ & $C_{\mathrm{det}}$ & Correct sign
& $A$ & $B$ & $1/B^2$ & $C_{\mathrm{det}}$ & Correct sign \\
\midrule
$e_{\mathrm{peak}}$ (GeV/fm$^3$) & 10 & 10.46 & 14.47 & 1.38 & 0.13 & 12/12 & 49.1 & 0.842 & 1.41 & $-0.30$ & 12/12 \\
$e_{\mathrm{peak}}$ (GeV/fm$^3$) & 20 & 10.00 & 13.65 & 1.37 & 0.54 & 11/12 & 40.9 & 0.844 & 1.41 & 0.50 & 11/12 \\
$e_{\mathrm{peak}}$ (GeV/fm$^3$) & 30 & 9.91 & 13.28 & 1.34 & 0.47 & 8/12 & 39.2 & 0.858 & 1.36 & 0.50 & 8/12 \\
$e_{\mathrm{peak}}$ (GeV/fm$^3$) & 40 & 9.84 & 13.21 & 1.34 & 0.54 & 11/12 & 37.7 & 0.866 & 1.33 & 0.33 & 11/12 \\
\addlinespace
$\sigma_\perp$ (fm) & 0.80 & 9.72 & 13.73 & 1.41 & 0.78 & 8/8 & 39.6 & 0.838 & 1.42 & 0.81 & 8/8 \\
$\sigma_\perp$ (fm) & 1.00 & 9.64 & 13.08 & 1.36 & 0.91 & 7/8 & 38.8 & 0.853 & 1.38 & 0.90 & 7/8 \\
$\sigma_\perp$ (fm) & 1.10 & 9.59 & 12.95 & 1.35 & 0.89 & 7/8 & 38.5 & 0.854 & 1.37 & 0.89 & 7/8 \\
$\sigma_\perp$ (fm) & 1.20 & 10.27 & 13.85 & 1.35 & 0.86 & 7/8 & 41.3 & 0.854 & 1.37 & 0.85 & 7/8 \\
$\sigma_\perp$ (fm) & 1.35 & 10.41 & 14.11 & 1.36 & 0.82 & 6/8 & 42.1 & 0.851 & 1.38 & 0.81 & 7/8 \\
$\sigma_\perp$ (fm) & 1.50 & 10.54 & 14.35 & 1.36 & 0.64 & 7/8 & 42.6 & 0.849 & 1.39 & 0.65 & 7/8 \\
\addlinespace
$b_0$ & 0.60 & 9.68 & 11.77 & 1.22 & 0.70 & 16/24 & 34.7 & 0.915 & 1.19 & 0.66 & 16/24 \\
$b_0$ & 0.77 & 10.11 & 15.27 & 1.51 & 0.71 & 19/24 & 45.8 & 0.799 & 1.57 & 0.70 & 18/24 \\
\bottomrule
\end{tabular}%
}
\end{table}

%% file: supplement_tables/prospective_states.tex
\begin{table}[H]
\caption{Eight additional states predicted with fixed coefficients. The
coefficients $A$ and $B$ of the compact relation, Eq.~(\ref{eq:total-sm}),
were fixed before these hydrodynamic calculations; the coefficients $a$ and
$b$ of the linear relation, Eq.~(\ref{eq:linear-sm}), were fitted later on the
48 calibration states only. Polarization harmonics are in units of
$10^{-3}$.}
\label{tab:prospective}
\centering
\resizebox{\textwidth}{!}{%
\begin{tabular}{lcccccccccc}
\toprule
State & Region & $b_0$ & $e_{\mathrm{peak}}$ & $\sigma_\perp$ & $\epsilon_{2,\mathrm{init}}$ & $\epsilon_{2,\mathrm{fo}}$ & $v_2$ & Direct & Eq.~(\ref{eq:linear-sm}) & Eq.~(\ref{eq:total-sm}) \\
 & & & (GeV/fm$^3$) & (fm) & & & & & & \\
\midrule
V01 & interpolation & 0.60 & 15 & 0.90 & 0.4378 & 0.3109 & 0.1783 & 0.920 & 0.680 & 0.629 \\
V02 & interpolation & 0.77 & 15 & 1.40 & 0.3571 & 0.3133 & 0.0985 & 1.727 & 1.792 & 1.803 \\
V03 & interpolation & 0.60 & 25 & 1.30 & 0.3629 & 0.2225 & 0.1493 & $-0.241$ & 0.191 & 0.141 \\
V04 & interpolation & 0.77 & 25 & 0.90 & 0.5132 & 0.2964 & 0.2184 & 0.389 & $-0.011$ & $-0.049$ \\
V05 & interpolation & 0.60 & 35 & 1.40 & 0.3448 & 0.1887 & 0.1422 & $-0.473$ & $-0.050$ & $-0.065$ \\
V06 & interpolation & 0.77 & 35 & 1.30 & 0.3846 & 0.2203 & 0.1490 & 0.210 & 0.173 & 0.116 \\
V07 & $\sigma_\perp$ extrapolation & 0.60 & 28 & 0.70 & 0.4734 & 0.2571 & 0.2255 & $-0.467$ & $-0.501$ & $-0.456$ \\
V08 & $\sigma_\perp$ extrapolation & 0.77 & 18 & 1.60 & 0.3081 & 0.2709 & 0.0879 & 1.667 & 1.512 & 1.537 \\
\bottomrule
\end{tabular}%
}
\end{table}

%% file: supplement_tables/b0_midpoint_states.tex
\begin{table}[H]
\caption{Four states at the previously unsampled reduced impact parameter
$b_0=0.685$, predicted with the coefficients of Eqs.~(\ref{eq:linear-sm}) and
(\ref{eq:total-sm}) fixed at their 48-state values. The endpoint midpoint is
the arithmetic mean of the direct total harmonics of the matching $b_0=0.60$
and $0.77$ states, and the departure is the direct total minus this midpoint.
Polarization harmonics and differences are in units of $10^{-3}$.}
\label{tab:b0-midpoint}
\centering
\resizebox{\textwidth}{!}{%
\begin{tabular}{lcccccccccc}
\toprule
State & $e_{\mathrm{peak}}$ & $\sigma_\perp$ & $\epsilon_{2,\mathrm{init}}$ & $\epsilon_{2,\mathrm{fo}}$ & $v_2$ & Direct & Eq.~(\ref{eq:linear-sm}) & Eq.~(\ref{eq:total-sm}) & Endpoint midpoint & Departure \\
 & (GeV/fm$^3$) & (fm) & & & & & & & & \\
\midrule
M01 & 10 & 0.80 & 0.5079 & 0.4151 & 0.1751 & 2.711 & 1.766 & 1.785 & 2.681 & $+0.031$ \\
M02 & 10 & 1.50 & 0.3415 & 0.3385 & 0.0785 & 1.897 & 2.317 & 2.406 & 1.682 & $+0.215$ \\
M03 & 30 & 0.80 & 0.5079 & 0.2606 & 0.2332 & $-0.294$ & $-0.571$ & $-0.489$ & $-0.369$ & $+0.075$ \\
M04 & 30 & 1.50 & 0.3415 & 0.2142 & 0.1326 & 0.172 & 0.336 & 0.271 & 0.255 & $-0.083$ \\
\bottomrule
\end{tabular}%
}
\end{table}

%% file: supplement_tables/model_checks.tex
\begin{table}[H]
\caption{Fixed-coefficient checks under changes of the decoupling temperature
$T_{\mathrm{dec}}$ and under the Gaussian pressure softening of
Eq.~(\ref{eq:pressure-softening}), at the initial state
$e_{\mathrm{peak}}=10\,\mathrm{GeV}/\mathrm{fm}^3$,
$\sigma_\perp=1.5\,\mathrm{fm}$, $b_0=0.77$. The standard bag-model result at
$T_{\mathrm{dec}}=160$ MeV is the common reference. Polarization harmonics are
in units of $10^{-3}$. Each error column gives the absolute difference between
the prediction and the direct total, divided by the mean magnitude of the two
cancelling contributions, $(|P_{z,s2}^{\omega}|+|P_{z,s2}^{\Xi}|)/2$, which
remains well defined near cancellation. Both relations give the correct sign
in every case.}
\label{tab:model-checks}
\centering
\resizebox{\textwidth}{!}{%
\begin{tabular}{lccccccc}
\toprule
& & & & \multicolumn{2}{c}{Linear, Eq.~(\ref{eq:linear-sm})}
& \multicolumn{2}{c}{Compact, Eq.~(\ref{eq:total-sm})} \\
\cmidrule(lr){5-6}\cmidrule(lr){7-8}
Variation & $\epsilon_{2,\mathrm{fo}}$ & $v_2$ & Direct & Prediction & Error & Prediction & Error \\
\midrule
$T_{\mathrm{dec}}=157.5$ MeV & 0.3411 & 0.0609 & 1.872 & 2.583 & 16\% & 2.561 & 15\% \\
$T_{\mathrm{dec}}=158.5$ MeV & 0.3421 & 0.0608 & 1.960 & 2.594 & 13\% & 2.576 & 13\% \\
$T_{\mathrm{dec}}=160.0$ MeV & 0.3468 & 0.0584 & 1.955 & 2.674 & 15\% & 2.644 & 15\% \\
$T_{\mathrm{dec}}=170.0$ MeV & 0.3724 & 0.0431 & 2.385 & 3.139 & 16\% & 2.931 & 11\% \\
Gaussian pressure softening & 0.3390 & 0.0455 & 1.675 & 2.772 & 44\% & 2.482 & 32\% \\
\bottomrule
\end{tabular}%
}
\end{table}

%% file: supplement_tables/shear_factored_comparison.tex
\begin{table}[H]
\caption{Comparison of the linear relation, Eq.~(\ref{eq:linear-sm}), the
compact vorticity-factored relation, Eq.~(\ref{eq:total-sm}), and the
kinematic-shear-factored relation, Eq.~(\ref{eq:shear-factored}). Each uses two
coefficients fitted to the same 48 calibration states. Each grouped
cross-validation refits both coefficients without the indicated group.
Coefficients remain fixed at their 48-state values for the eight additional
states and the four states at $b_0=0.685$; the last row combines these 12
states. The root-mean-square error (RMSE) is in units of $10^{-3}$.}
\label{tab:shear-comparison}
\centering
\resizebox{\textwidth}{!}{%
\begin{tabular}{lccccccccc}
\toprule
& \multicolumn{3}{c}{Linear, Eq.~(\ref{eq:linear-sm})}
& \multicolumn{3}{c}{Compact, Eq.~(\ref{eq:total-sm})}
& \multicolumn{3}{c}{Kinematic-shear-factored, Eq.~(\ref{eq:shear-factored})} \\
\cmidrule(lr){2-4}\cmidrule(lr){5-7}\cmidrule(lr){8-10}
Test & $C_{\mathrm{det}}$ & RMSE & Correct sign
& $C_{\mathrm{det}}$ & RMSE & Correct sign
& $C_{\mathrm{det}}$ & RMSE & Correct sign \\
\midrule
48-state fit & 0.866 & 0.409 & 41/48 & 0.865 & 0.410 & 43/48 & 0.836 & 0.453 & 41/48 \\
Leave $e_{\mathrm{peak}}$ group out & 0.860 & 0.418 & 42/48 & 0.813 & 0.483 & 42/48 & 0.760 & 0.548 & 42/48 \\
Leave $\sigma_\perp$ group out & 0.834 & 0.454 & 42/48 & 0.838 & 0.450 & 43/48 & 0.797 & 0.503 & 41/48 \\
Leave $b_0$ group out & 0.740 & 0.569 & 35/48 & 0.724 & 0.587 & 34/48 & 0.693 & 0.619 & 33/48 \\
Eight additional states & 0.890 & 0.277 & 6/8 & 0.888 & 0.279 & 6/8 & 0.880 & 0.288 & 6/8 \\
Four states at $b_0=0.685$ & 0.806 & 0.541 & 4/4 & 0.807 & 0.540 & 4/4 & 0.758 & 0.605 & 4/4 \\
All 12 additional states & 0.860 & 0.386 & 10/12 & 0.860 & 0.386 & 10/12 & 0.833 & 0.421 & 10/12 \\
\bottomrule
\end{tabular}%
}
\end{table}

%% file: supplement_tables/shear_factored_inversion.tex
\begin{table}[H]
\caption{Freeze-out eccentricity $\epsilon_{2,\mathrm{fo}}$ inferred from the
central values of the ALICE longitudinal-polarization harmonics at $5.02$ and
$5.36\,\mathrm{TeV}$, with the spectrum-weighted $\Lambda$ elliptic flow $v_2$
at $5.02\,\mathrm{TeV}$ as input at both energies (main text, Sec.~VI A). The
$5.36\,\mathrm{TeV}$ polarization values were extracted from the published
vector figure. The interval for Eq.~(\ref{eq:total-sm}) varies the initial
eccentricity $\epsilon_{2,\mathrm{init}}$ over the envelope used in the main
text; Eqs.~(\ref{eq:linear-sm}) and (\ref{eq:shear-factored}) contain no
$\epsilon_{2,\mathrm{init}}$ and give single values. Experimental
uncertainties and the $\pm5\%$ variation of the $v_2$ input are not included.}
\label{tab:shear-inversion}
\centering
\begin{tabular}{ccccc}
\toprule
$\sqrt{s_{\mathrm{NN}}}$ (TeV) & Centrality & Eq.~(\ref{eq:linear-sm}) & Eq.~(\ref{eq:total-sm}) & Eq.~(\ref{eq:shear-factored}) \\
\midrule
5.02 & 30--40\% & 0.242 & 0.245--0.270 & 0.250 \\
5.02 & 40--50\% & 0.276 & 0.271--0.306 & 0.286 \\
5.36 & 30--40\% & 0.208 & 0.216--0.233 & 0.215 \\
5.36 & 40--50\% & 0.274 & 0.270--0.305 & 0.284 \\
\bottomrule
\end{tabular}
\end{table}

%% file: supplement_tables/local_calibration.tex
\begin{table}[H]
\caption{Local calibration of Eq.~(\ref{eq:linear-sm}) for the ALICE
inversion. Each row is a fit over six PICR states at the reduced impact
parameter matched to the centrality class, with
$e_{\mathrm{peak}}=20$ and $40\,\mathrm{GeV}/\mathrm{fm}^3$ (multiplied by
$2^{4/3}$ for $t_0=2\,\mathrm{fm}/c$) and $\sigma_\perp=1.5$, $1.8$, and
$2.1\,\mathrm{fm}$. The coefficients $a$ and $b$ are in units of $10^{-3}$,
with standard errors from the fit; their correlation coefficient is about
$0.97$ in every row. $\langle p_T\rangle$ is the $\Lambda$ mean transverse
momentum in $0.6<p_T<6\,\mathrm{GeV}/c$ (GeV/$c$); ALICE measures $1.58$ and
$1.53$ in 30--40\% and 40--50\% at $5.02\,\mathrm{TeV}$. The inverted $\efo$
uses the central values of the $5.36\,\mathrm{TeV}$ inputs of the main text; the interval
is the 68\% range including the experimental uncertainties, the $v_2$-proxy
variation, and the coefficient uncertainty; the last column splits the
central value into $\Pz/a+(b/a)v_2$.}
\label{tab:local-calib}
\centering
\resizebox{\textwidth}{!}{%
\begin{tabular}{llccccccccc}
\toprule
Centrality & $t_0$ (fm/$c$) & $a$ & $b$ & $b/a$ & $C_{\mathrm{det}}$ & $v_2$ & $\efo$ & $\langle p_T\rangle$
& inverted $\efo$ (5.36 TeV) & $\Pz/a+(b/a)v_2$ \\
\midrule
30--40\% ($b_0=0.60$) & 4 & $6.3\pm1.5$ & $10.3\pm2.7$ & 1.63 & 0.83 & 0.075--0.136 & 0.153--0.231 & 1.07--1.30 & 0.27 (0.24--0.30) & $0.073+0.193$ \\
30--40\% ($b_0=0.60$) & 2 & $13.8\pm2.3$ & $18.7\pm3.0$ & 1.35 & 0.90 & 0.089--0.147 & 0.127--0.190 & 1.13--1.42 & 0.19 (0.18--0.21) & $0.033+0.160$ \\
40--50\% ($b_0=0.685$) & 4 & $8.0\pm1.3$ & $12.6\pm2.5$ & 1.59 & 0.98 & 0.067--0.136 & 0.161--0.252 & 1.04--1.27 & 0.33 (0.30--0.36) & $0.123+0.205$ \\
40--50\% ($b_0=0.685$) & 2 & $11.1\pm1.8$ & $14.0\pm2.5$ & 1.26 & 0.99 & 0.082--0.154 & 0.138--0.203 & 1.10--1.38 & 0.25 (0.23--0.27) & $0.088+0.163$ \\
\bottomrule
\end{tabular}%
}
\end{table}

%% file: supplement_tables/blastwave_alice.tex
\begin{table}[H]
\caption{Blast-wave background taken from the ALICE blast-wave fit to
$\pi^\pm$, $K^\pm$, and (anti)proton spectra in Pb-Pb collisions at
$5.02\,\mathrm{TeV}$~\cite{alicebwsm}: mean transverse velocity
$\langle\beta_T\rangle$, kinetic freeze-out temperature $T_{\mathrm{kin}}$,
and profile exponent $n$, with the surface velocity
$\beta_s=\langle\beta_T\rangle(n+2)/2$. The ranges cover the source radius
$r_0=8$, $10$, $12\,\mathrm{fm}$ and the time depth $\Delta\tau=2$, $3$, $4$,
$6\,\mathrm{fm}$ at $T_f=T_{\mathrm{kin}}$, with the central decoupling proper
time fixed at $9\,\mathrm{fm}/c$; $a$ and $b$ are in units of
$10^{-3}$. $\langle p_T\rangle$ is the $\Lambda$ mean transverse momentum in
$0.6<p_T<6\,\mathrm{GeV}/c$ (GeV/$c$), the window of the measured values $1.58$
and $1.53$. The inverted $\efo$ is given as
minimum, median, and maximum over the twelve backgrounds for the central values of the inputs.}
\label{tab:bw-alice}
\centering
\resizebox{\textwidth}{!}{%
\begin{tabular}{lcccccccccc}
\toprule
Centrality & $\langle\beta_T\rangle$ & $T_{\mathrm{kin}}$ (GeV) & $n$ & $\beta_s$ & $a$ & $b$ & $b/a$
& $\langle p_T\rangle$ & $\efo$ (5.36 TeV) & $\efo$ (5.02 TeV) \\
\midrule
30--40\% & 0.622 & 0.101 & 0.83 & 0.880 & 5.4--16.5 & 8.9--20.9 & 1.27--1.65 & 1.66--1.74 & 0.18, 0.23, 0.28 & 0.20, 0.27, 0.34 \\
40--50\% & 0.595 & 0.108 & 0.91 & 0.865 & 5.4--15.3 & 8.6--19.2 & 1.26--1.59 & 1.61--1.68 & 0.23, 0.30, 0.39 & 0.23, 0.30, 0.39 \\
\bottomrule
\end{tabular}%
}
\end{table}

%% file: supplement_tables/pion_hbt.tex
\begin{table}[H]
\caption{Pion femtoscopic eccentricity on 35 PICR surfaces (positive flux).
$r$ is the correlation with $\efo$; $C_{\mathrm{det}}$ and signs refer to
Eq.~(\ref{eq:linear-sm}) refitted with $\epsilon_{\mathrm{HBT}}$ in place of
$\efo$.}
\label{tab:hbt}
\centering
\begin{tabular}{lcccc}
\toprule
$k_T$ (GeV/$c$) & $\langle\epsilon_{\mathrm{HBT}}/\efo\rangle$ & $r$ & $C_{\mathrm{det}}$ & signs\\
\midrule
$\to0$ & 0.92 & 0.80 & 0.72 & 29/35\\
$0.10$ & 0.91 & 0.77 & 0.67 & 29/35\\
$0.25$ & 0.91 & 0.70 & 0.53 & 28/35\\
$0.35$ & 0.92 & 0.65 & 0.38 & 26/35\\
$0.45$ & 0.95 & 0.58 & 0.22 & 25/35\\
$0.60$ & 1.02 & 0.48 & 0.03 & 20/35\\
average $0.25$--$0.60$ & 0.95 & 0.60 & 0.27 & 25/35\\
$\efo$ (reference) & -- & -- & 0.75 & 30/35\\
\bottomrule
\end{tabular}
\end{table}

%% file: supplement_tables/training_states.tex
\footnotesize
\setlength{\tabcolsep}{5.0pt}
\begin{longtable}{rcccccc}
\caption{Initial-state, freeze-out-geometry, and elliptic-flow values for the complete 48-state calibration set.}\label{tab:training-states}\\
\toprule
State & $b_0$ & $e_{\mathrm{peak}}$ (GeV/fm$^3$) & $\sigma_\perp$ (fm) & $\epsilon_{2,\mathrm{init}}$ & $\epsilon_{2,\mathrm{fo}}$ & $v_2$ \\
\midrule
\endfirsthead
\multicolumn{7}{c}{Table~\ref{tab:training-states} continued}\\
\toprule
State & $b_0$ & $e_{\mathrm{peak}}$ (GeV/fm$^3$) & $\sigma_\perp$ (fm) & $\epsilon_{2,\mathrm{init}}$ & $\epsilon_{2,\mathrm{fo}}$ & $v_2$ \\
\midrule
\endhead
\midrule
\multicolumn{7}{r}{Continued on next page}\\
\endfoot
\bottomrule
\endlastfoot
1 & 0.60 & 10 & 0.80 & 0.4560 & 0.3721 & 0.1597 \\
2 & 0.60 & 10 & 1.00 & 0.4191 & 0.3545 & 0.1402 \\
3 & 0.60 & 10 & 1.10 & 0.4002 & 0.3457 & 0.1298 \\
4 & 0.60 & 10 & 1.20 & 0.3814 & 0.3402 & 0.1145 \\
5 & 0.60 & 10 & 1.35 & 0.3537 & 0.3221 & 0.0998 \\
6 & 0.60 & 10 & 1.50 & 0.3272 & 0.3076 & 0.0857 \\
7 & 0.77 & 10 & 0.80 & 0.5487 & 0.4672 & 0.1827 \\
8 & 0.77 & 10 & 1.00 & 0.4786 & 0.4250 & 0.1401 \\
9 & 0.77 & 10 & 1.10 & 0.4455 & 0.4106 & 0.1177 \\
10 & 0.77 & 10 & 1.20 & 0.4140 & 0.3943 & 0.0974 \\
11 & 0.77 & 10 & 1.35 & 0.3705 & 0.3718 & 0.0755 \\
12 & 0.77 & 10 & 1.50 & 0.3316 & 0.3456 & 0.0544 \\
13 & 0.60 & 20 & 0.80 & 0.4560 & 0.2863 & 0.2029 \\
14 & 0.60 & 20 & 1.00 & 0.4191 & 0.2700 & 0.1773 \\
15 & 0.60 & 20 & 1.10 & 0.4002 & 0.2614 & 0.1627 \\
16 & 0.60 & 20 & 1.20 & 0.3814 & 0.2542 & 0.1520 \\
17 & 0.60 & 20 & 1.35 & 0.3537 & 0.2462 & 0.1368 \\
18 & 0.60 & 20 & 1.50 & 0.3272 & 0.2306 & 0.1241 \\
19 & 0.77 & 20 & 0.80 & 0.5487 & 0.3458 & 0.2349 \\
20 & 0.77 & 20 & 1.00 & 0.4786 & 0.3171 & 0.1918 \\
21 & 0.77 & 20 & 1.10 & 0.4455 & 0.3046 & 0.1737 \\
22 & 0.77 & 20 & 1.20 & 0.4140 & 0.2952 & 0.1487 \\
23 & 0.77 & 20 & 1.35 & 0.3705 & 0.2864 & 0.1248 \\
24 & 0.77 & 20 & 1.50 & 0.3316 & 0.2675 & 0.1058 \\
25 & 0.60 & 30 & 0.80 & 0.4560 & 0.2384 & 0.2095 \\
26 & 0.60 & 30 & 1.00 & 0.4191 & 0.2260 & 0.1844 \\
27 & 0.60 & 30 & 1.10 & 0.4002 & 0.2201 & 0.1711 \\
28 & 0.60 & 30 & 1.20 & 0.3814 & 0.2107 & 0.1597 \\
29 & 0.60 & 30 & 1.35 & 0.3537 & 0.2032 & 0.1481 \\
30 & 0.60 & 30 & 1.50 & 0.3272 & 0.1934 & 0.1304 \\
31 & 0.77 & 30 & 0.80 & 0.5487 & 0.2787 & 0.2484 \\
32 & 0.77 & 30 & 1.00 & 0.4786 & 0.2645 & 0.2047 \\
33 & 0.77 & 30 & 1.10 & 0.4455 & 0.2526 & 0.1832 \\
34 & 0.77 & 30 & 1.20 & 0.4140 & 0.2462 & 0.1642 \\
35 & 0.77 & 30 & 1.35 & 0.3705 & 0.2338 & 0.1425 \\
36 & 0.77 & 30 & 1.50 & 0.3316 & 0.2269 & 0.1204 \\
37 & 0.60 & 40 & 0.80 & 0.4560 & 0.2076 & 0.2144 \\
38 & 0.60 & 40 & 1.00 & 0.4191 & 0.1937 & 0.1812 \\
39 & 0.60 & 40 & 1.10 & 0.4002 & 0.1871 & 0.1751 \\
40 & 0.60 & 40 & 1.20 & 0.3814 & 0.1858 & 0.1590 \\
41 & 0.60 & 40 & 1.35 & 0.3537 & 0.1803 & 0.1453 \\
42 & 0.60 & 40 & 1.50 & 0.3272 & 0.1693 & 0.1358 \\
43 & 0.77 & 40 & 0.80 & 0.5487 & 0.2409 & 0.2482 \\
44 & 0.77 & 40 & 1.00 & 0.4786 & 0.2284 & 0.2063 \\
45 & 0.77 & 40 & 1.10 & 0.4455 & 0.2249 & 0.1840 \\
46 & 0.77 & 40 & 1.20 & 0.4140 & 0.2145 & 0.1712 \\
47 & 0.77 & 40 & 1.35 & 0.3705 & 0.2038 & 0.1508 \\
48 & 0.77 & 40 & 1.50 & 0.3316 & 0.1925 & 0.1242 \\
\end{longtable}
\clearpage
\footnotesize
\setlength{\tabcolsep}{8.0pt}
\begin{longtable}{rccccc}
\caption{Polarization harmonics of the complete 48-state calibration set and the fitted total harmonics of the linear relation, Eq.~(\ref{eq:linear-sm}), and of the compact relation, Eq.~(\ref{eq:total-sm}). The total ILE harmonic is $P_{z,s2}^{\mathrm{ILE}}=P_{z,s2}^{\omega}+P_{z,s2}^{\Xi}$. All numerical columns are in units of $10^{-3}$.}\label{tab:training-harmonics}\\
\toprule
State & $P_{z,s2}^{\omega}$ & $P_{z,s2}^{\Xi}$ & $P_{z,s2}^{\mathrm{ILE}}$ & Eq.~(\ref{eq:linear-sm}) & Eq.~(\ref{eq:total-sm}) \\
\midrule
\endfirsthead
\multicolumn{6}{c}{Table~\ref{tab:training-harmonics} continued}\\
\toprule
State & $P_{z,s2}^{\omega}$ & $P_{z,s2}^{\Xi}$ & $P_{z,s2}^{\mathrm{ILE}}$ & Eq.~(\ref{eq:linear-sm}) & Eq.~(\ref{eq:total-sm}) \\
\midrule
\endhead
\midrule
\multicolumn{6}{r}{Continued on next page}\\
\endfoot
\bottomrule
\endlastfoot
1 & -6.239 & 8.103 & 1.863 & 1.546 & 1.567 \\
2 & -5.033 & 6.494 & 1.461 & 1.636 & 1.686 \\
3 & -4.639 & 6.612 & 1.973 & 1.690 & 1.755 \\
4 & -4.749 & 6.421 & 1.673 & 1.843 & 1.920 \\
5 & -3.645 & 5.303 & 1.658 & 1.862 & 1.936 \\
6 & -3.764 & 5.045 & 1.281 & 1.909 & 1.988 \\
7 & -6.257 & 9.755 & 3.498 & 2.184 & 2.259 \\
8 & -5.199 & 8.319 & 3.120 & 2.343 & 2.416 \\
9 & -4.340 & 7.233 & 2.894 & 2.504 & 2.576 \\
10 & -4.122 & 6.220 & 2.098 & 2.617 & 2.663 \\
11 & -3.948 & 5.962 & 2.014 & 2.691 & 2.713 \\
12 & -2.721 & 4.804 & 2.083 & 2.716 & 2.618 \\
13 & -5.271 & 5.075 & -0.196 & 0.099 & 0.052 \\
14 & -4.482 & 4.733 & 0.251 & 0.285 & 0.227 \\
15 & -3.940 & 4.034 & 0.094 & 0.398 & 0.334 \\
16 & -3.861 & 4.011 & 0.150 & 0.471 & 0.407 \\
17 & -4.017 & 4.241 & 0.224 & 0.599 & 0.541 \\
18 & -3.737 & 3.830 & 0.093 & 0.616 & 0.562 \\
19 & -5.780 & 6.576 & 0.796 & 0.258 & 0.189 \\
20 & -5.612 & 6.450 & 0.838 & 0.558 & 0.479 \\
21 & -3.813 & 5.149 & 1.336 & 0.680 & 0.605 \\
22 & -4.209 & 5.095 & 0.886 & 0.927 & 0.848 \\
23 & -3.901 & 5.086 & 1.185 & 1.164 & 1.123 \\
24 & -3.784 & 5.210 & 1.427 & 1.234 & 1.214 \\
25 & -4.558 & 4.155 & -0.403 & -0.470 & -0.411 \\
26 & -3.695 & 3.268 & -0.427 & -0.252 & -0.237 \\
27 & -3.967 & 3.327 & -0.640 & -0.130 & -0.136 \\
28 & -3.825 & 3.312 & -0.513 & -0.069 & -0.084 \\
29 & -3.629 & 3.321 & -0.307 & 0.014 & -0.014 \\
30 & -3.368 & 3.001 & -0.367 & 0.158 & 0.110 \\
31 & -4.544 & 4.210 & -0.334 & -0.597 & -0.507 \\
32 & -4.742 & 4.850 & 0.108 & -0.144 & -0.153 \\
33 & -3.809 & 3.804 & -0.005 & 0.030 & -0.007 \\
34 & -4.005 & 4.158 & 0.154 & 0.225 & 0.162 \\
35 & -3.517 & 3.955 & 0.438 & 0.397 & 0.323 \\
36 & -3.120 & 3.998 & 0.878 & 0.629 & 0.557 \\
37 & -3.592 & 2.356 & -1.235 & -0.845 & -0.644 \\
38 & -3.747 & 3.002 & -0.745 & -0.532 & -0.409 \\
39 & -3.386 & 2.450 & -0.936 & -0.515 & -0.400 \\
40 & -3.480 & 2.848 & -0.632 & -0.308 & -0.253 \\
41 & -3.645 & 3.285 & -0.360 & -0.177 & -0.158 \\
42 & -3.268 & 3.132 & -0.135 & -0.157 & -0.144 \\
43 & -4.780 & 4.438 & -0.342 & -0.973 & -0.715 \\
44 & -5.333 & 4.475 & -0.859 & -0.527 & -0.419 \\
45 & -2.915 & 2.648 & -0.267 & -0.258 & -0.225 \\
46 & -3.399 & 3.468 & 0.068 & -0.187 & -0.173 \\
47 & -3.962 & 4.032 & 0.070 & -0.016 & -0.039 \\
48 & -2.682 & 3.130 & 0.448 & 0.233 & 0.170 \\
\end{longtable}